\documentclass[useAMS,usenatbib]{mn2e}

\usepackage{amsmath}
\usepackage{amsfonts}
\usepackage{amssymb}
\usepackage{graphicx}
\usepackage{epstopdf}
\usepackage{epsfig}
\DeclareGraphicsExtensions{.pdf,.png,.jpg}
\graphicspath{{./}}
\usepackage{lscape}
\usepackage{bold-extra}
\usepackage{appendix}
\usepackage{listings}
\usepackage{float}
\usepackage{pdfpages}
\usepackage{natbib}

\def\micron{$\mu$m }
\def\Halpha{H$\alpha$ }
\usepackage{url}
\usepackage{hyperref}
\usepackage{enumerate}
\usepackage{setspace}
\let\oldsqrt\sqrt
\def\sqrt{\mathpalette\DHLhksqrt}
\def\DHLhksqrt#1#2{%
\setbox0=\hbox{$#1\oldsqrt{#2\,}$}\dimen0=\ht0
\advance\dimen0-0.2\ht0
\setbox2=\hbox{\vrule height\ht0 depth -\dimen0}%
{\box0\lower0.4pt\box2}}
\title[z=2.5 protocluster associated with HzRG MRC 2104-242]{A z=2.5 protocluster associated with the radio galaxy MRC 2104-242: star formation and differing mass functions in dense environments}
\author[E.A.Cooke et al.]{E. A. Cooke$^{1}$\thanks{e-mail: ppxeaco@nottingham.ac.uk}, N. A.
Hatch$^{1}$, S. I. Muldrew$^{1}$, E. E. Rigby$^{2}$ and J. D. Kurk$^{3}$\\
$^{1}$School of Physics and Astronomy, University of Nottingham, University Park, Nottingham NG7 2RD\\
$^{2}$Leiden Observatory, P.O. Box 9513, 2300 RA, Leiden, The Netherlands\\
$^{3}$Max-Planck-Institut fuer Extraterrestrische Physik, Giessenbachstrasse, D-85748 Garching, Germany}
\begin{document}

\date{Accepted 2014 March 13. Received 2014 March 10; in original form 2013 December 13}

\pagerange{\pageref{firstpage}--\pageref{lastpage}} \pubyear{2014}

\maketitle

\label{firstpage}

\begin{abstract}
We present results from a narrow-band survey of the field around the high redshift radio galaxy MRC\,2104$-$242. We have selected \Halpha emitters in a 7\,sq.\,arcmin field and compared the measured number density with that of a field sample at similar redshift. We find that MRC\,2104$-$242 lies in an overdensity of galaxies that is $8.0 \pm 0.8$ times the average density of a blank field, suggesting it resides in a large-scale structure that may eventually collapse to form a massive cluster. We find that there is more dust obscured star formation in the protocluster galaxies than in similarly selected control field galaxies and there is tentative evidence of a higher fraction of starbursting galaxies in the denser environment. However, on average we do not find a difference between the star formation rate (SFR)-mass relations of the protocluster and field galaxies and so conclude that the SFR of these galaxies at $z\sim2.5$ is governed predominantly by galaxy mass and not the host environment. We also find that the stellar mass distribution of the protocluster galaxies is skewed towards higher masses and there is a significant lack of galaxies at $M < 10^{10} M_{\odot}$ within our small field of view. Based on the level of overdensity we expect to find $\sim22$ star forming galaxies below $10^{10} M_{\odot}$ in the protocluster and do not detect any. This lack of low mass galaxies affects the level of overdensity which we detect. If we only consider high mass ($M > 10^{10.5} M_{\odot}$) galaxies, the density of the protocluster field increases to $\sim55$ times the control field density. 
\end{abstract}


\section{Introduction}
Locally, the star formation rate (SFR)-mass relation does not change as a function of galaxy environment; the fraction of galaxies which are star forming differs but the specific star formation rate (sSFR) is constant irrespective of environment \citep{YPeng2010}. This SFR-mass relation evolves with redshift, however cluster and field galaxies continue to lie on the same relation up to $z=1$ \citep{Muzzin2012}. At higher redshifts, studies have found that this trend of a constant sSFR between galaxies in the process of forming a cluster (protocluster galaxies) and field galaxies appears to continue, implying a sSFR independent of environment \citep{Koyama2013a,Koyama2013b}. The existence of a ``main sequence" for galaxies suggests that star formation in galaxies proceeds in the same way in (proto)clusters as it does in the field, even at redshifts $z>2$.  Protocluster galaxy properties, however, differ from those in the field: the progenitors of low redshift clusters have previously been found to contain member galaxies that are older, more star-forming, more metal-rich and twice as massive as field galaxies at the same redshift \citep{Steidel2005,Hatch2011b,Koyama2013a,Kulas2013}. This implies that cluster galaxies have experienced an accelerated growth in their early years, yet their sSFRs show no difference from the field up to redshift $z = 2$.

Previously, the SFR-mass relation at $z>2$ has been studied using masses derived from K-band fluxes, and SFRs corrected using mass-dependent dust extinction estimates \citep{Koyama2013a,Koyama2013b}. Using a dust extinction law that is solely dependent on the mass of the object makes it difficult to find extreme starbursts that lie above the main sequence. Using the rest frame UV slope as a direct measure of dust extinction, as well as infrared star formation indicators such as 24\,\micron and 250\,\micron fluxes, may help to break this degeneracy between normal star-forming galaxies and heavily dust-obscured star-bursting objects. Combining this with SED-derived masses should provide a better measure of the SFR-mass relation for protocluster and field galaxies at $z >2$.

In this paper we investigate the SFR-mass relation in a candidate protocluster field, around the radio galaxy MRC\,2104$-$242. This field was observed as part of an infrared survey of eight high-redshift radio galaxies (HzRGs), described in \citet{Galametz2010b} and \citet{Hatch2011a}. Four of these HzRGs appeared to be surrounded by an overdensity of red galaxies, one of which (MRC\,0156$-$252) has recently been spectroscopically confirmed to lie within a large-scale structure \citep{Galametz2013b}. 
Another of these targets, MRC\,2104$-$242, had a 3$\sigma$ overdensity %  (density of field/control field density $-1$) 
of red galaxies and the angular correlation function showed that the galaxies in this field were more clustered than average \citep{Hatch2011a}. 
MRC\,2104$-$242 lies at $z = 2.49$, which means the \Halpha emission line falls directly within the ISAAC narrow-band filter at 2.29\,$\mu$m. This allows us to select star-forming galaxies within a narrow redshift range ($\Delta z = 0.05$) around the radio galaxy. Using optical to MIR photometry we have studied the masses and star-forming properties of \Halpha selected galaxies around MRC\,2104$-$242. We have compared the results in the radio galaxy field to a control field sample, using the same selection techniques throughout.

The outline of the paper is as follows: Section \ref{sec:data} outlines the observations, data reduction and sample selection. Section \ref{sec:properties} describes our methods in determining the galaxy properties. In Section \ref{sec:results} we present our results and look at galaxy properties as a function of environment. Section \ref{sec:discussion} discusses our key results and possible implications and Section \ref{sec:summary} presents a summary. We assume a $\Lambda$CDM cosmology with H$_0 = 70$\,km\,s$^{-1}$\,Mpc$^{-1}$, $\Omega_M = 0.3$ and $\Omega_\Lambda = 0.7$ throughout, unless stated otherwise. We adopt a \citet{Chabrier2003} initial mass function (IMF) for all our calculations and magnitudes are given in the AB system unless stated otherwise.

\section{Data} \label{sec:data}
MRC\,2104$-$242 lies at a redshift of 2.49 \citep{McCarthy1990} and has been found to lie in an overdensity of red galaxies \citep[$J-H > H-K + 0.5$ \, $\cap$ \, $J-K > 1.5$, see][]{Hatch2011a}. We have obtained photometry of this target in $g'$, $z'$, $J$, $H$, $K_s$,  3.6\,$\mu$m, 4.5\,$\mu$m, and 24\,$\mu$m bands as well as narrow-band photometry at 2.29\,$\mu$m, covering an area of 2.65\,arcmin $\times$ 2.65\,arcmin. This narrow-band filter is centred on the \Halpha emission line at $z=2.49$, the redshift of the radio galaxy. The width of the filter (324\,\AA) allows us to select \Halpha emitters between $2.46 < z < 2.51$. This corresponds to $\Delta v \sim 4300$\,km\,s$^{-1}$, so we expect to detect all protocluster members. 

\subsection{Imaging and data reduction}

\subsubsection{NIR observations}
MRC\,2104$-$242 was observed in service mode using the High Acuity Wide-field K-band Imager (HAWK-I) \citep{Kissler-Patig2008} to obtain the H, J and K$_\text{s}$ images, and ISAAC to obtain the narrow-band (hereafter NB) 2.29\,\micron image. Details on the observations and reduction of the H, J and K$_\text{s}$ data are provided in \citet{Hatch2011a}. The NB data were obtained in 2011 October 8-10th for a total integration time of 5.6\,h. The ISAAC field of view is smaller than the HAWK-I field of view (2.5\,arcmin $\times$ 2.5\,arcmin compared to 7.5\,arcmin $\times$ 7.5\,arcmin), so the detector was aligned to match the coverage of the HAWK-I chip containing the radio galaxy. The radio galaxy was positioned in the upper-right section of the ISAAC detector to match the spatial coverage of the deep HAWK-I data.

The NB data were reduced with the ESO/MVM data reduction pipeline \citep{Vandame2004} and the astrometric solutions were derived using a catalogue from the K$_\text{s}$ HAWK-I data. The pixel scale of the H, J and K$_\text{s}$ HAWK-I images (0.106\,arcsec\,pixel$^{-1}$) was degraded to the ISAAC pixel scale of 0.148\,arcsec\,pixel$^{-1}$. The NB image was convolved to the seeing of the K$_\text{s}$ of 0.7\,arcsec.

The total overlapping area of the NB, H, J and K$_\text{s}$ images is 11.8\,sq.\,arcmin, resulting from the large dithering pattern used during the NB observations. To ensure the image depth was approximately consistent across the whole image, regions which had less than 30 percent of the maximum exposure time were masked out. The remaining area is $7.09$\,sq.\,arcmin. The 3$\sigma$ image depths given in Table \ref{table:imagedepths} were measured by placing $2$\,arcsec apertures at multiple ($\sim 10000$) random locations.  

The NB image was flux-calibrated using the HAWK-I K$_\text{s}$ image (which was flux-calibrated using 2MASS stars in the field of view; see \citealt{Hatch2011a}) and further adjustments were made to this calibration by comparing the $NB - K_s$ colour of stars in the images to the predicted colours of stars in the Pickles stellar library. Uncertainties in the flux calibration are $<$0.04\,mag. No correction was applied to account for Galactic extinction as this is negligible.

\subsubsection{MIR and FIR observations}
IRAC \citep{Fazio2004} observations at 3.6\,\micron and 4.5\,\micron were obtained in 2009 during the warm \emph{Spitzer} mission (PID 60112) for a total integration time of 1600\,s in both bands. Details of the observations and data reduction can be found in \citet{Galametz2012}. The limiting magnitudes for the IRAC bands were estimated from their completeness curves. 

\emph{Spitzer} MIPS \citep{Rieke2004} 24\,\micron data was obtained as part of the \emph{Spitzer} High-redshift Radio Galaxy sample survey. Full details of the observations and data reduction can be found in \citet{Seymour2007}.

\emph{Herschel} SPIRE \citep{Griffin2010} 250\,\micron imaging was obtained during the Search for Protoclusters with \emph{Herschel} (SPHer) survey. The depth of the SPIRE data of the MRC\,2104$-$242 field is identical to that of the three control fields. A description of the data can be found in \citet{Rigby2014}. 

\begin{table} %\begin{table*}
\begin{centering}
% \resizebox{17cm}{!}{
\begin{tabular}{ l c c c }
\hline
Filter & Integration Time & 3$\sigma$ Limit (AB) & Instrument \\ \hline \hline
$g'$ & 3.8\,h & 27.8 & GMOS-S \\
$z'$ & 0.67\,h & 25.1 & GMOS-S \\
$J$ & 3.38\,h & 25.3 & HAWK-I \\
$H$ & 0.67\,h & 24.3 & HAWK-I \\
$K_s$ & 1.53\,h & 24.0 & HAWK-I \\
NB229 & 5.6\,h & 21.4 & ISAAC \\ 
3.6\,\micron & 0.44\,h & 23.0 & IRAC \\
4.5\,\micron & 0.44\,h & 22.7 & IRAC \\ \hline
\end{tabular}
% }
\caption{Details of the images used. Limiting magnitudes for the optical and NIR images were measured using randomly placed 2\,arcsec apertures. The IRAC image limits were determined from their completeness curves.}
\label{table:imagedepths}
\end{centering}
\end{table} %\end{table*}

% \subsection{Data reduction}
% $g'$ and $z$ band from GMOS \\
\subsubsection{Optical observations}
Observations in the optical regime ($g'$ and $z'$ bands) were taken in service mode using the Gemini Multi-Object Spectrograph South \citep[GMOS-S;][]{Hook2004} instrument on Cerro Pachon, Chile, during the period August--November 2010.  The $z'$ band total integration time was 40\,min, % taken in 8 exposures of 180\,s.
and the total $g'$ band integration time was 3.8\,h%, taken in 8 exposures of 180\,s and 41 exposures of 300\,s
. The $g'$ and $z'$ data were reduced using the Gemini {\sc gemtools} IRAF package. The usual reduction steps were taken: bias subtraction, flat fielding, and trimming of the image. The $z'$ band fringing was removed using IDL to subtract the fringe frame, which had been created using the IRAF package {\sc gifringe}. The images were mosaiced and combined using {\sc imcombine}.

The $g'$ image was flux-calibrated by comparing the $g'-J$ colour of stars in the image to those predicted using the Pickles stellar library (the $J$ image was flux-calibrated using 2MASS stars in the field of view; see \citealt{Hatch2011a}). The $z'$ image was then flux-calibrated similarly, using the $g'-z'$ colour of stars. 3$\sigma$ image depths were measured by placing $\sim 10000$ random 2\,arcsec apertures on the images.

\begin{figure}
\centering
\includegraphics[scale=0.5]{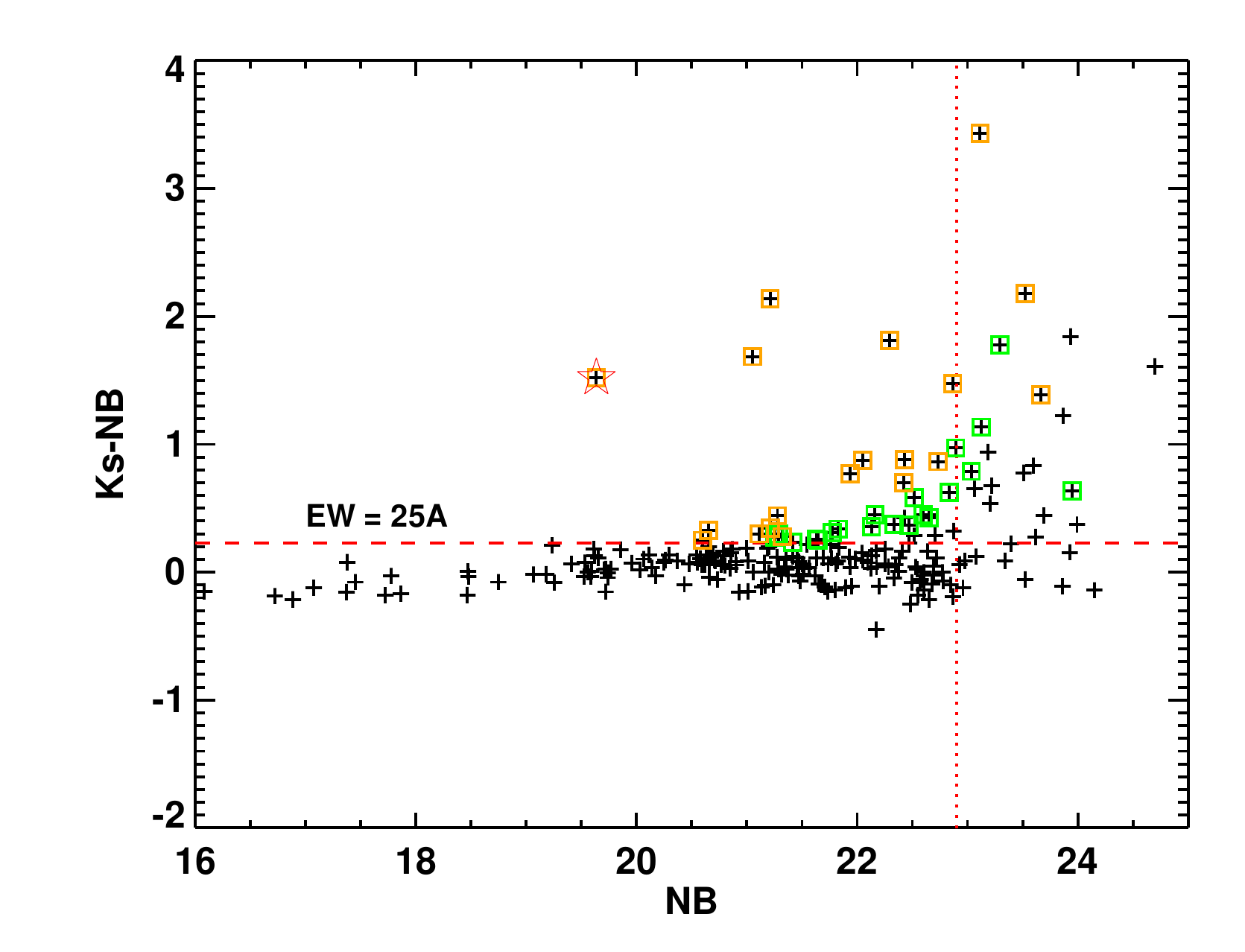}
\caption{Colour-magnitude diagram for the MRC\,2104$-$242 field. Green and orange points highlight NB-excess sources with $K_s-NB > 2\Sigma$ and $K_s-NB > 3\Sigma$, respectively. The radio galaxy is highlighted with a red star. The dashed line marks a rest-frame EW cut of $25 \text{\AA}$. The vertical dotted line shows the 80\% completeness limit in the NB.
}
\label{fig:colmag}
\end{figure}

\subsection{Control field}
We compare our radio galaxy field to three control fields taken from the Ultra Deep Survey (UDS), the Cosmic Evolution Survey (COSMOS) and the Great Observatories Origins Deep Survey-South (GOODS-S). We have photometry in approximately the same bands as our radio galaxy field ($B$, $z'$, $J$, $H$, $K_s$, 3.6\,$\mu$m, 4.5\,$\mu$m, 24\,$\mu$m, 250\,$\mu$m). NB images were taken using the HAWK-I H$_2$ 2.12\,\micron filter for the UDS and COSMOS fields and using the NB2090 filter for the GOODS-S field. These filters detect \Halpha emission at $2.22 \leq z \leq 2.26$ and $2.18 \leq z \leq 2.21$ respectively. When calculating densities we scale our control field results according to the different volumes given by each filter. 
Each of our control fields is limited by the size of the NB field-of-view and are all approximately $57$\,sq.\,arcmin. We refer to \citet{Hatch2011b} for details on the reduction of the K$_\text{s}$ and NB images. The remaining photometry was obtained from public archives and is described in \citet{Capak2007,Capak2011,Furusawa2008,Retzlaff2010,McCracken2012,Hartley2013}. The \emph{Spitzer} data was obtained from the NASA/IPAC Infrared Science Archive. The \emph{Herschel} 250\,\micron data was obtained from the H-ATLAS survey \citep{H-ATLAS} and re-reduced to have the same depth as the MRC\,2104$-$242 data, see \citet{Rigby2014} for details. 

\subsection{Catalogues}

The {\sc SExtractor} software package \citep{Bertin1996} was used to create a photometric catalogue of our data. We used {\sc SExtractor} in dual-image mode, using a weighted NB image as the detection image, to obtain fluxes in all bands. The NB image was weighted with the square root of the effective exposure map, which takes background noise into account. We select as sources those with 25 adjoining pixels that are 1$\sigma$ above the rms background and use apertures of 2\,arcsec in diameter for measuring colours. These apertures are significantly larger than the $\sim0.7$\,arcsec FWHM of point sources in the images.  

Individual flux densities were measured using Kron {\sc auto} apertures.  
Limiting magnitudes for the optical and NIR bands were estimated by measuring the standard deviation of the flux densities in 2\,arcsec diameter apertures placed randomly on the images (Table \ref{table:imagedepths}).
For the IRAC 3.6\,\micron and 4.5\,\micron bands, {\sc SExtractor} was optimised with $\text{{\sc minarea}} = 4$ pixels and $\text{{\sc detect\_thresh}} = 2.5 \sigma$ above the rms background. The NB photometric catalogues were matched with the IRAC catalogues within 1\,arcsec using {\sc Topcat} \citep{Topcat} to produce the full photometric catalogue. 
In order to determine what effect the choice of {\sc SExtractor} parameters had on our results, we checked our methods using three different parameter combinations: 2\,arcsec fixed apertures (25 adjoining pixels), {\sc auto} apertures for 25 adjoining pixels and {\sc auto} apertures with 24 adjoining pixels. We found that the choice of selection parameters does not significantly affect our results and does not alter our conclusions.

\subsection{Selection of NB sources}
To obtain a sample of NB-excess sources we followed the method of \citet{Bunker1995}, selecting sources with excess NB signal relative to the $K_s$ band. Sources with a value of $K_s-NB \geq 2\Sigma$ were selected as NB excess sources, with $\Sigma$ defined as: 

\begin{equation} \label{eq:sigma}
\Sigma = \frac{1-10^{-0.4(K-NB)}}{10^{-0.4(zp-NB)}\sqrt{\pi r^2_{ap}(\sigma^2_{NB}+\sigma^2_{K})}}
\end{equation}

$K$ and NB are the AB magnitudes in each band, $\sigma$ values are the {\sc SExtractor} errors for each band, $\pi r_{ap}^2$ is the area of the aperture used and $zp$ is the zero-point of the images; here $zp = 26.9$.

\begin{figure*}
\centering
\includegraphics[width=2\columnwidth]{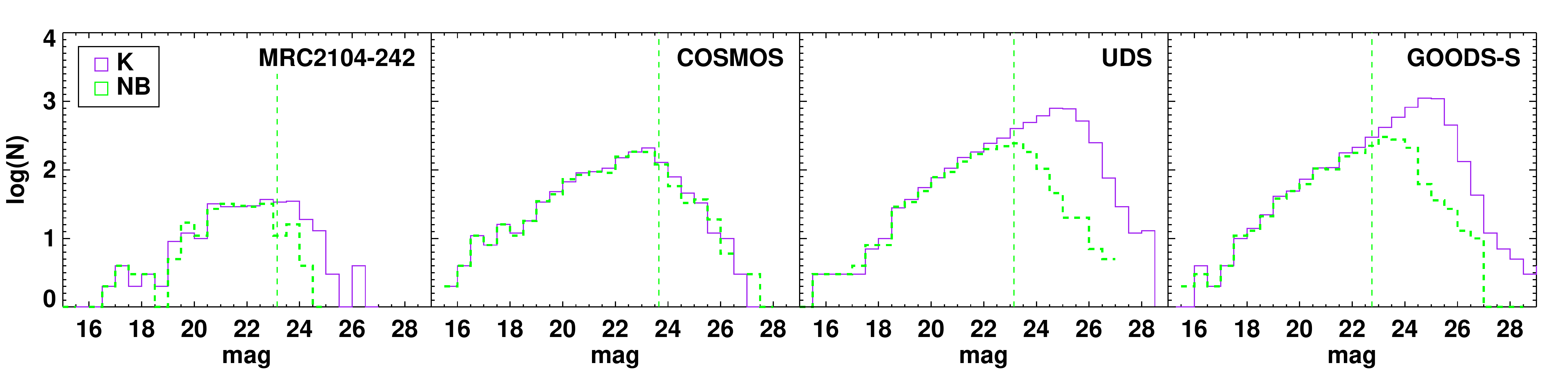}
\caption{Completeness histograms for the $K_s$ (purple lines) and NB (green dashed lines) images in the MRC\,2104$-$242 field and control fields. Vertical lines mark the 80\% completeness limit for each NB image.
}
\label{fig:completeness}
\end{figure*}

A rest frame equivalent width (EW) cut of $25$\,$\text{\AA}$ was also used to avoid contamination due to photometric errors. 
In Figure \ref{fig:colmag} we plot the $K_s-NB$ colours against the NB magnitudes for all sources. $\Sigma$ quantifies the significance of the NB excess and our 2$\Sigma$ selection corresponds to a completeness cut in star formation rate (SFR) of $\sim 7$ M$_{\odot}$ yr$^{-1}$. We also exclude sources with NB magnitude fainter than $22.9$. At this limit we are $>80$\% complete in both the radio galaxy field and all the control fields. Completeness was calculated by comparing the detection catalogues for the NB and deeper $K_s$ images. In Figure \ref{fig:completeness} we plot the completeness curves for each field in the NB and $K_s$ band. Vertical lines indicate where the NB becomes 80\% complete. Our $NB >22.9$\,mag corresponds to the completeness of the radio galaxy field. In the radio galaxy field we find 31 sources above this limit, 16 of which have values of $K_s-NB > 3\Sigma$. 14 of these NB excess sources have detections at 3.6 and 4.5\,$\mu$m.

\subsection{\Halpha emitters}
Excess NB flux could also be produced from low-redshift ($z < 1$) emission line contaminants or [O{\scshape{iii}}] lines from sources at $z = 3.57$.

To remove low-redshift contaminants we used two methods: firstly following the method of \citet{Daddi2004}, we select \Halpha emitters as sources with BzK colours \\ ($(z-K_s)-(B-z) > -0.2$, or equivalently gzK colours: \\ $(z-K_s)-\left(\frac{(g'-z')-0.13}{0.87}\right) > -0.2$). The BzK criterion selects sources that lie at redshifts between $1.4 < z < 2.5$ and has a contamination rate of $\leq 13\%$ from galaxies at $z<1$ \citep{Daddi2004}. We do not have $B$ band photometry in the radio galaxy field so we used the $g'$ band photometry in its place. We converted the selection criteria using model galaxy spectra, redshifted to the lower limit of BzK-selected galaxies ($z = 1.4$) and convolved with $B$, $g'$ and $z'$ filters. A line was fit to the $g'-z'$ versus $B-z'$ points to obtain the selection conversion.   
Secondly, for sources with IRAC detections, a colour cut of $[3.6]-[4.5] > -0.1$ was taken, selecting sources which lie at $z > 1.3$ \citep{Papovich2008}. 

We retain in our sample those sources which are selected by either the BzK or IRAC criterion. 
We remove two sources because they appear to be associated with a large, foreground galaxy, possibly a spiral. We have checked our results with and without including these sources and they remain unchanged. We therefore remove the sources to avoid contamination from low redshift interlopers. 

\citet{Sobral2013} find that 10-20\% of sources selected using the BzK method may be high redshift contaminants. However, without spectroscopic information we are unable to identify sources at $z = 3.57$ and cannot remove them from our sample. 
After applying our selection to our NB excess sources, we have 18 \Halpha emitters in our sample (from 31 NB excess sources), including the radio galaxy and three ``companion'' galaxies, which lie within 3\,arcsec of the radio galaxy. 9 of these \Halpha emitters were selected via the IRAC colour selection, and 11 via the BzK criterion (2 were selected by both criteria). We select $17/25$, $9/16$, $8/12$ (\Halpha emitters $/$ NB excess sources) from the COSMOS, UDS and GOODS-S control fields respectively.

\subsection{AGN} \label{sec:AGN}
We estimate the contamination rate of AGN in our control fields using the \emph{Spitzer} IRAC criterion from \citet{Donley2012}. From this selection we estimate that there are two possible AGN in the COSMOS \Halpha emitter sample and none in the UDS or GOODS-S samples. We do not have 5.8\,\micron and 8\,\micron data for the MRC\,2104$-$242 field that is deep enough to determine the number of AGN around the radio galaxy. Assuming the AGN fraction in the MRC\,2104$-$242 field is the same as in the control fields (AGN$/$\Halpha emitters $= 0.03$)%(0.013\,AGN\,sq.arcmin$^{-1}$)
, we do not expect to find any AGN in this field. We leave the suspected AGN in our control field sample, so we do not bias our results, but discuss how removing them will affect our results in Section \ref{sec:robust_agn}.

\section{Determining properties of \Halpha emitters} \label{sec:properties}
\subsection{Stellar mass}
We determined stellar masses by using the SED fitting programme ``Fitting and Assessment of Synthetic Templates" \citep[FAST,][]{Kriek2009a} to fit the photometry of our sample of \Halpha candidates to obtain mass estimates. We assume from now on that the NB excess flux in the \Halpha candidates is due to H$\alpha+$[N{\scshape{ii}}] emission at %a redshift $2.46 \leq z \leq 2.51$. 
the redshift of the radio galaxy and we fixed the redshift of the fit to $z = 2.49$%, corresponding to the width of the NB filter
. The control field galaxy redshifts were set to $z=2.24,2.24,$ and $2.19$ for COSMOS, UDS and GOODS-S respectively, assuming \Halpha emission from the centre of the NB filters. 

We used FAST to fit \citet{BC03} stellar population synthesis models with a \citet{Chabrier2003} IMF to our photometry ($B$/$g'$,$z'$,$J$,$H$,$K$,$[3.6]$,$[4.5]$). $12/18$ \Halpha emitters in the MRC\,2104$-$242 field had detections in the IRAC bands. We fit delayed exponentially declining (SFR $\sim t \exp[-t/\tau]$) star formation histories with dust extinction $0 < A_V < 3$ in steps of 0.2\,mag (assuming the \citet{Calzetti2000} extinction law), $7.0 < \text{log}_{10}(\tau/\text{yr}) < 10.1$ in steps of 0.1 and $7.5 < \text{log}_{10}(\text{age}/\text{yr}) < 9.5$ in steps of 0.2. Metallicities were fixed to solar abundance. 
As we have rest-frame UV, optical and NIR photometry, the stellar mass output from the SED is well-determined. Due to degeneracies between SFR, dust extinction ($A_V$) and the assumed star formation histories, we do not use these outputs from the FAST output as they are likely to be highly unreliable. However, the mass output is robust independent of the exact star formation history template that is assumed \citep{Shapley2005}. 
Errors in the stellar masses are determined from 100 Monte Carlo simulations performed by FAST, with the photometry being varied within the flux uncertainties. We also added a rest-frame template error function to take into account the uncertainties in the model templates. 

Some of the photometry for the control fields is deeper than for the protocluster field. In our analysis only detections to the depth of the MRC\,2104$-$242 field were considered in the control fields. 
We have checked our results using full-depth magnitudes for the control field and find that our overall conclusions are unaffected by the different depths of the images between fields.

\begin{figure}
\centering
%Trim option's parameter order: left bottom right top
\includegraphics[width=\columnwidth]{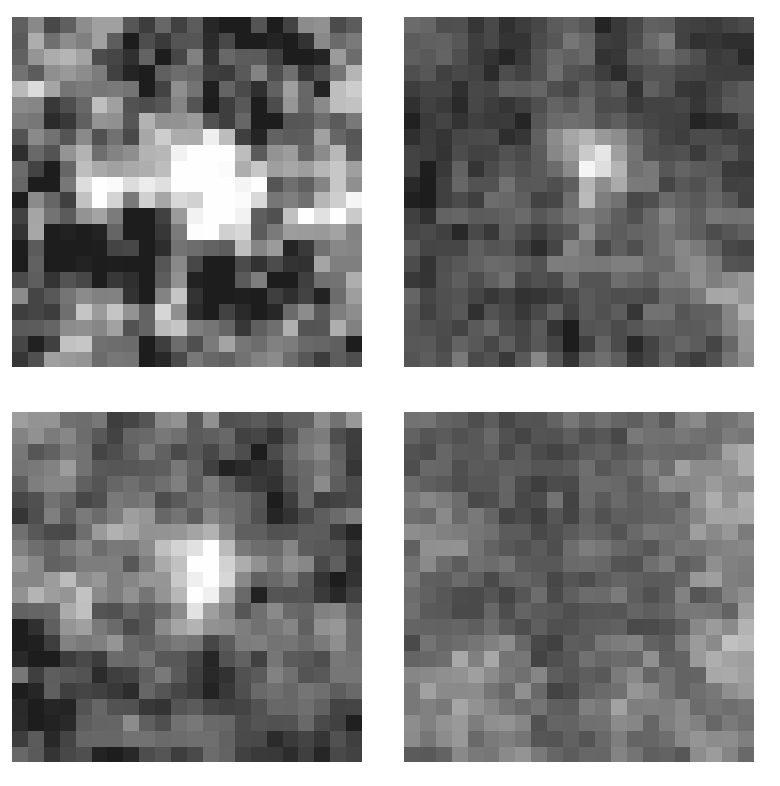}
\caption{Median stacks of MIPS 24\,\micron images for \Halpha emitters. Clockwise from top left: MRC\,2104$-$242 (14 stamps), COSMOS (17 stamps), GOODS-S (8 stamps), UDS (9 stamps). All images have the same scale. Three of the four fields have clear detections, with MRC\,2104$-$242 showing a stronger signal. The radio galaxy and companions are not included in the stack, however the COSMOS AGN candidates are included.}
\label{fig:MIPSstacks}
\end{figure}

\subsection{SFRs}
\subsubsection{H$\alpha$-derived SFRs}
We calculate the $K_s$ continuum and convert our NB signal to an \Halpha flux using: 

\begin{equation} \label{eq:f(Kcont)}
f(K_{\text{cont}}) = \frac{w_{K_s}f(K_s)-w_{\text{NB}}f(\text{NB})}{w_{K_s}-w_{\text{NB}}}
\end{equation}

\begin{equation} \label{eq:f(Ha)}
f(\text{H}\alpha) = w_{\text{NB}}[f(\text{NB})-f(K_{\text{cont}})]
\end{equation}

where $f(K_{\text{cont}})$ is the continuum flux density in the $K_s$ band, $f(\text{NB})$ and $f(K_s)$ are the flux densities in the NB and $K_s$ bands respectively, $f(\text{H}\alpha)$ is the \Halpha flux, and $w_{K_s}$ and $w_{\text{NB}}$ are the widths of the corresponding filters.

These values are corrected for dust extinction calculated from the $B-z'$ colour\footnote{For the MRC\,2104$-$242 field the $B-z'$ colour was calculated using $(B-z') = \left(\frac{(g'-z')+0.09}{0.91}\right)$ at $z = 2.5$}, which corresponds to the rest-frame UV slope, following the method of \citet{Daddi2004}: 

\begin{equation}
E(B-V) = 0.25(B-z' + 0.1)_{AB}
\end{equation}

Note that here we assume that the extinction for \Halpha is the same as for the broadband SED. Where sources had $g'$, $B$ or $z'$ magnitudes fainter than the $3\sigma$ limiting magnitude (see Table \ref{table:imagedepths}) we convolved the best fitting SED template for that source with the appropriate filter curve in order to get a magnitude estimate. 
For the radio galaxy field any sources with $g'$ magnitudes fainter than $3$ times the limiting magnitude were convolved with a $B$ filter curve to avoid having to convert the colours. For each of the control fields and for the radio galaxy field $z'$ band, we used the $B$ or $z'$ filter curve of the instrument used to obtain the data.

Dust-corrected \Halpha luminosities were then calculated, scaling for luminosity distance, and \Halpha SFRs determined using the \citet{Kennicutt1998} relation, converted to a \citet{Chabrier2003} IMF:

\begin{equation} \label{eq:Kennicutt}
SFR (\text{M}_{\odot}\text{\, yr}^{-1}) = 4.39 \times 10^{-42} L_{H\alpha} (\text{erg\,s}^{-1})
\end{equation}

\begin{figure*}
\centering
%Trim option's parameter order: left bottom right top
\includegraphics[width=2\columnwidth]{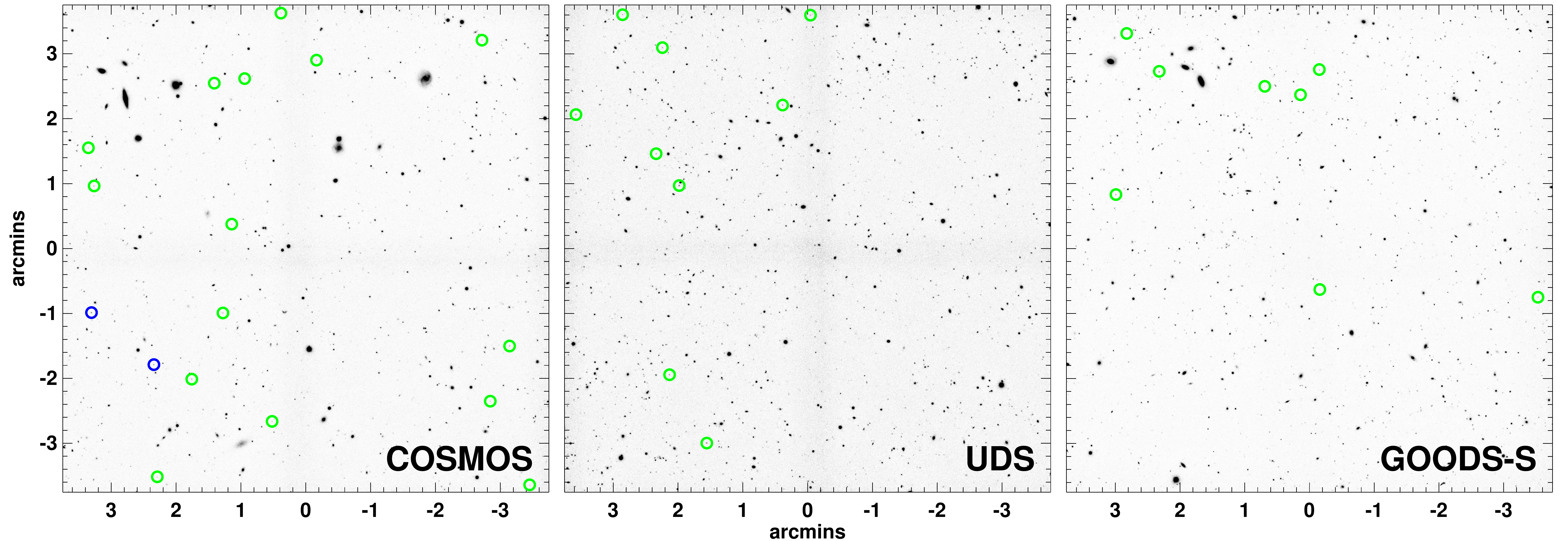}
\caption{The control fields used in this study. From left: COSMOS, UDS, GOODS-S. The figures show the NB images, with detected \Halpha sources overlaid as green circles. The AGN candidates in the COSMOS field are highlighted in blue. Each window is 7.5\,arcmin $\times$ 7.5\,arcmin.}
\label{fig:overdensity1}
\end{figure*}

\subsubsection{MIPS 24\,\micron SFRs} \label{sec:MIPS}
The \emph{Spitzer} 24\,\micron filter transmits between $20.8$-$25.8 \mu$m, which corresponds to rest-frame wavelengths of $6.0$-$7.4$\,\micron for $z=2.49$ galaxies. This rest-frame wavelength range is dominated by polycyclic aromatic hydrocarbon (PAH) features, which have been shown to provide a good measure of hidden star formation \citep{Siana2009}. 

The 24\,\micron data have a $3\sigma$ detection limit of $\sim 0.11$\,mJy. We have a $>3\sigma$ detection in 24\,\micron for the radio galaxy and its companions (these sources are blended in the 24\,\micron image), however the majority of our \Halpha emitters were not individually detected. We therefore stacked the sources to obtain a median flux density for each field. The radio galaxy and its companions were not included in the stack, however we include the AGN candidates in the COSMOS field as these sources were not individually detected at $>2\sigma$. Postage stamps of 22 $\times$ 22 pixels ($4.5$ times the \emph{Spitzer} 24\,\micron FWHM) were created around each \Halpha source, and sources in each field were median stacked (Figure \ref{fig:MIPSstacks}). Flux densities were then measured from the stacks in 8 pixel (5\,arcsec) diameter apertures (Table \ref{table:MIPSflux}). 
These rest-frame IR flux densities were converted to SFRs using both the methods outlined in \citet{Rujopakarn2013} (their section 5) and using equation 14 of \citet{Rieke2009}\footnote{%\begin{equation} %\label{eq:SFR_IR}
$\log{\left(SFR_{IR}\right)} = 0.108 + 1.711(\log{\left(4 \pi L_{d}^{2} f\right)} - 53)$
%  = 4.3 \times 10^{-10} \frac{4 \pi L_{d}^{2}}{L_{\odot}} f
%\end{equation}
%%NB- Hartley2013 uses Rujopakarn2011

Here, $f$ is the flux density in an 8 pixel diameter aperture, $L_d$ is the luminosity distance in cm.}. The method from \citet{Rujopakarn2013} assumes these galaxies lie on the galaxy main sequence (MS), whereas \citet{Rieke2009} calculate the SFR for (ultra) luminous infrared galaxies ([U]LIRGs). Without additional information, such as a measure of the IR bump, we cannot distinguish between the two scenarios for the galaxies in our sample \citep[see][]{Elbaz2011} and so use both methods in our analysis.

The detection limit of $0.11$\,mJy corresponds to $\sim 145$\,M$_{\odot}$\,yr$^{-1}$ or $\sim 1200$\,M$_{\odot}$\,yr$^{-1}$ (MS or ULIRG) at $z = 2.5$.

\begin{figure}
\centering
%Trim option's parameter order: left bottom right top
\includegraphics[width=0.9\columnwidth]{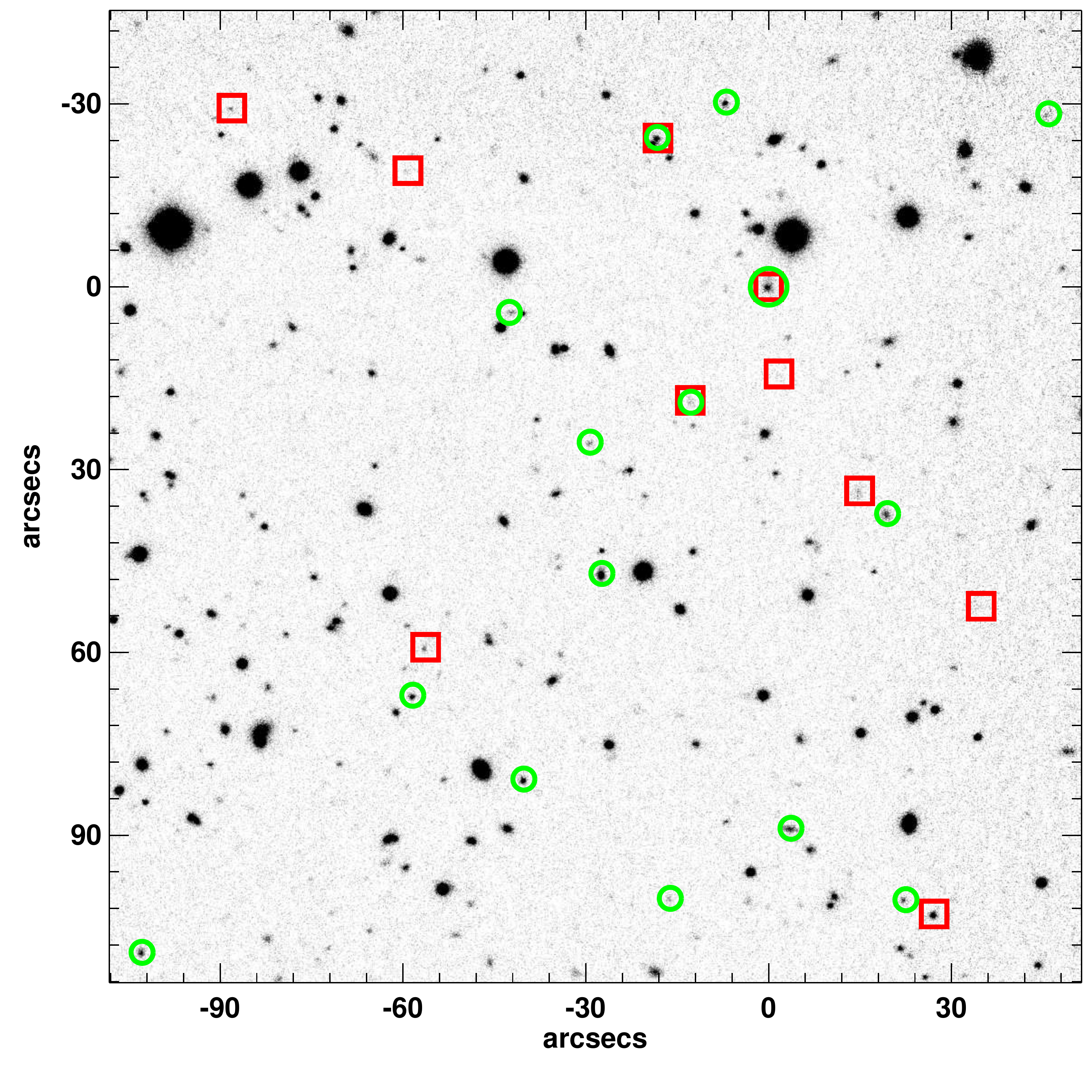}
\caption{K$_{\text{s}}$ image of the field around MRC\,2104$-$242. North is up, East to the left. Detected \Halpha sources are shown with green circles. The radio galaxy and three companions (see Figure \ref{fig:zoom}) lie at the origin, within the larger green circle of radius 3\,arcsec. The window size is 2.65\,arcmin $\times$ 2.65\,arcmin. The MRC\,2104$-$242 field is clearly overdense compared to the control fields (see also Figure \ref{fig:overdensity1}), containing 14 \Halpha emitters in a $\sim 7$\,sq. arcmin field. For comparison we also show galaxies selected by the JHK criterion (red squares, see text for details). The radio galaxy was also selected by the JHK criterion.}
\label{fig:overdensity2}
\end{figure}

\begin{figure}
\centering
%Trim option's parameter order: left bottom right top
\includegraphics[width=0.9\columnwidth]{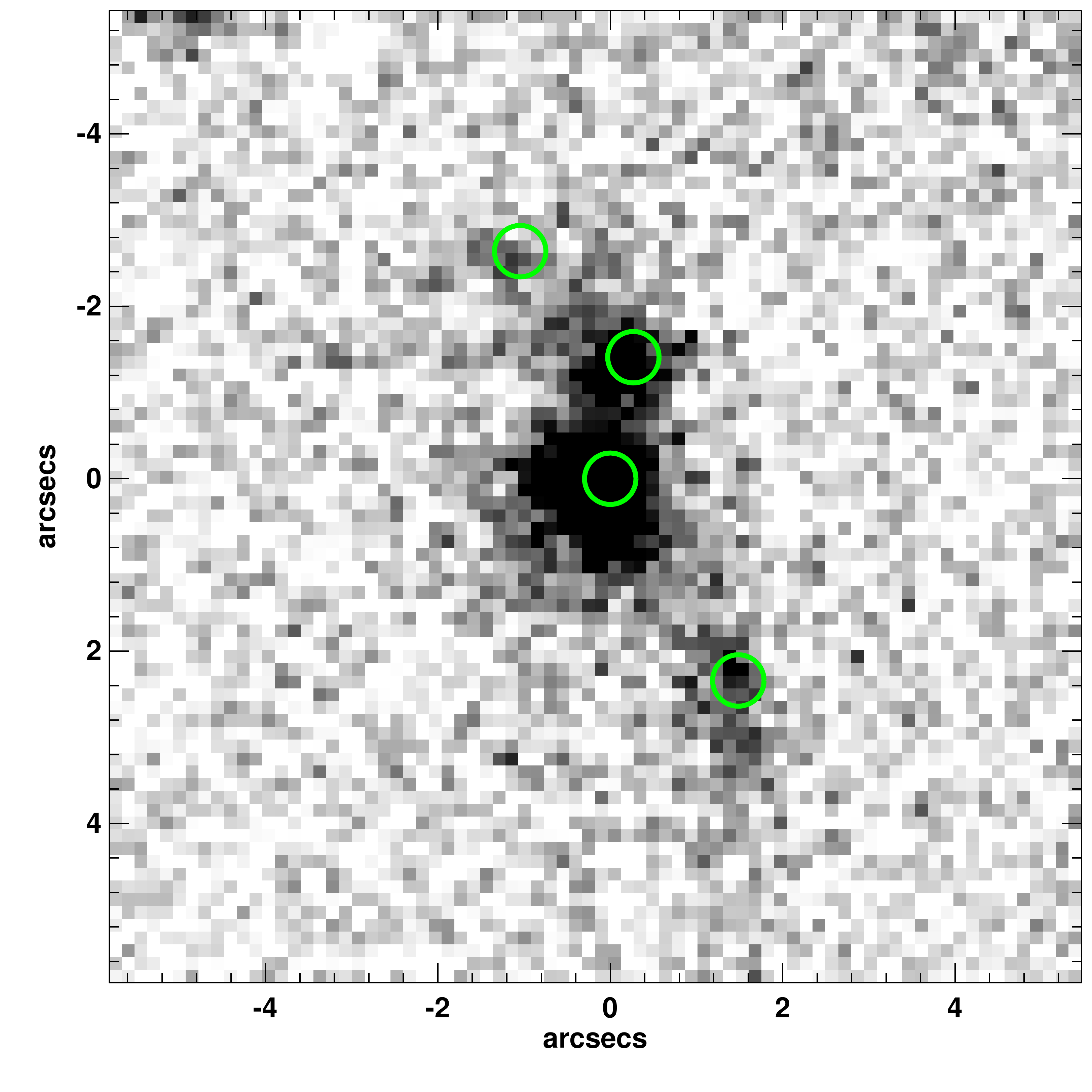}
\caption{NB image of the radio galaxy MRC\,2104$-$242 and its three companion sources, all circled in green.}
\label{fig:zoom}
\end{figure}

\subsubsection{\emph{Herschel} 250\,\micron SFRs} 
The \emph{Herschel} SPIRE 250\,\micron filter probes the far-IR bump for galaxies at $z > 2$, allowing the total IR luminosity of distant galaxies to be measured. 
These data have a $3\sigma$ detection limit of $\sim 375$\,M$_{\odot}$\,yr$^{-1}$ at $z = 2.5$. The radio galaxy and its companions are detected in the \emph{Herschel} 250\,\micron data, and a few other \Halpha sources had $>2\sigma$ detections within 10\,arcsec, however due to the large beam size of \emph{Herschel} we are unable to robustly identify counterparts. To obtain an estimate of the SFR of the \Halpha emitters we therefore median stacked all \Halpha sources (not including the radio galaxy and its companions).   
A SFR was derived from the median 250\,\micron flux by modelling the IR bump as an isothermal body of temperature 35\,Kelvin and $\beta=1.5$. This template was normalised to the detected 250\,\micron flux and integrated over 8-1000\,\micron to obtain L$_{IR}$. The L$_{IR}$ was converted to a SFR using the \citet{Kennicutt1998} relation adjusted to a Chabier IMF by dividing the SFRs by $1.6$. Median stacks of the \Halpha emitters in the UDS, COSMOS and GOODS-S fields were produced in the same manner, but none of these stacks resulted in a signal above 3$\sigma$ significance. %Each stack contains 9-17 \Halpha sources. Assuming Poisson noise, we would require $> 20$ galaxies per stack in order to detect an average SFR of $\sim 40$\,M$_{\odot}$\,yr$^{-1}$ at a three sigma level. 

\begin{table*}
\begin{centering}
\begin{tabular}{ l l l l l l }
\hline
Field & $n$ & Flux density ($\mu$Jy) & SFR (MS; M$_{\odot}$\,yr$^{-1}$) & SFR (ULIRG; M$_{\odot}$\,yr$^{-1}$) \\ \hline \hline
MRC\,2104$-$242 & 14 & 35.7 $\pm$ 10.0 & $37.3 \pm 13.3$ & $171.4 \pm 94.6$ \\
COSMOS & 17 & 10.3 (9.2) $\pm$ 3.5 (3.8)$^{a}$ & $6.3 \pm 2.6$ & $13.3 \pm 8.5$ \\
UDS & 9 & 18.1 $\pm$ 1.2 & $12.3 \pm 1.4$ & $34.7 \pm 5.9$ \\
GOODS-S & 8 & -$^{b}$ $\pm$ 0.52 & 0.63 & 0.48 \\ \hline
\end{tabular}
\caption{Flux densities measured from the 24\micron stacks in an aperture of radius 5\,arcsec. The uncertainties are the standard deviation of 1000 sets of $n$ stacked random regions (where $n$ is the number of \Halpha sources in each field). The SFRs given are calculated from the 24\,\micron fluxes using relations based on local ULIRGs and main sequence (MS) estimates. $^{a}$ Numbers in brackets for COSMOS are flux density and error values when the AGN candidates are removed from the stack. $^{b}$ There was no detectable signal in the GOODS-S stack, we use the 3$\sigma$ value in all SFR calculations.}
\label{table:MIPSflux}
\end{centering}
\end{table*}

\section{Results} \label{sec:results}
\subsection{Galaxy overdensity} \label{sec:overdensity}
The field around MRC\,2104$-$242 has a large overdensity of \Halpha emitters (Figures \ref{fig:overdensity1} \& \ref{fig:overdensity2}). Excluding the radio galaxy and three nearby companions there are 14 objects in a 7.09\,sq.\,arcmin field, which is $8.0 \pm 0.8$ times the density of the control fields, i.e. contains a galaxy overdensity of $7.0 \pm 0.8$. The field of view around the HzRG is relatively small ($4.5 \text{\,Mpc} \times 4.5 \text{\,Mpc}$ comoving) compared to the average size of high redshift protoclusters: protoclusters at $z > 2$ typically extend for $\sim 10$\,Mpc \citep{Venemans2007,Hatch2011a}. As \citet{Chiang2013} show this means we cannot say anything for certain about the mass of this structure, however, this level of overdensity is of the same order that has been found in other protoclusters at similar redshift \citep[e.g.][]{Kurk2004a,Hatch2011b,Hayashi2012}. MRC\,2104$-$242 is therefore likely to also lie within a protocluster. % which will collapse into a massive ($> 10^{14} M_{\odot}$) cluster by redshift $z=0$.

We tested to see if there was any preferential clustering of \Halpha sources around the radio galaxy. We did this by comparing the average distance from the radio galaxy to average distances calculated from random distributions of sources. The average distance of the \Halpha sources from the radio galaxy differs from that expected from a random distribution at a 2.6\,sigma level. However, this includes the three companion galaxies within 3\,arcsec of the radio galaxy. When  these three sources are excluded from the analysis the significance is only 1.2\,sigma. Therefore there is no strong clustering around the radio galaxy.

\subsection{Red galaxies}
\begin{figure}
\centering
\includegraphics[width=\columnwidth]{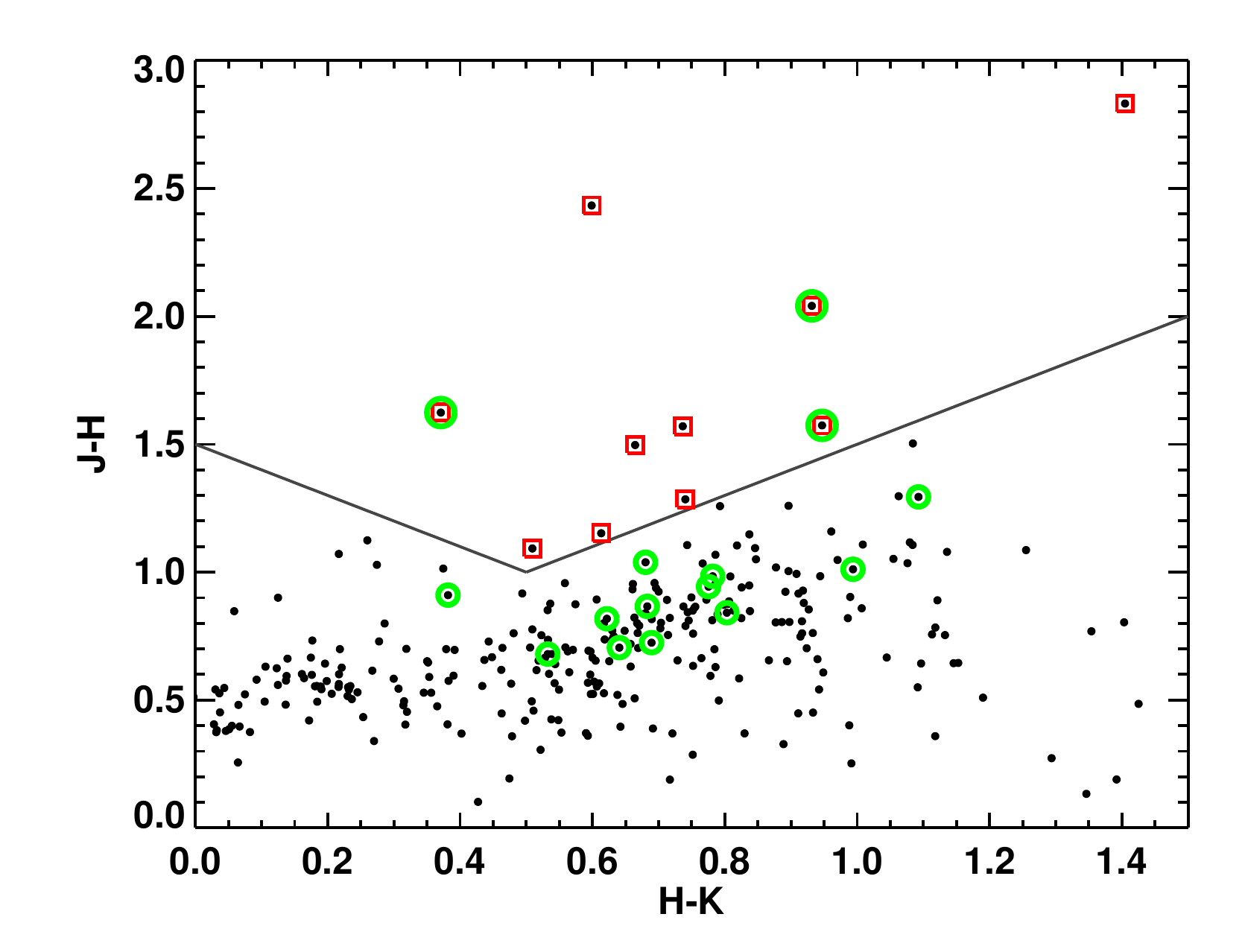}
\caption{Near IR colours of galaxies in the MRC\,2104$-$242 field. Lines mark the JHK criterion used to select red galaxies at high redshift; galaxies selected this way are shown by red squares. \Halpha emitters are highlighted with green circles.}
\label{fig:JHK}
\end{figure}

\citet{Hatch2011a} found a 3\,$\sigma$ overdensity of JHK galaxies  ($J-H > H-K + 0.5$ \, $\cap$ \, $J-K > 1.5$ [Vega]) around MRC\,2104$-$242. The JHK criterion selects red galaxies with low SFRs or star forming galaxies which are heavily obscured by dust, and so probes a different population to the \Halpha emitters. We find 10 JHK galaxies within the ISAAC field-of-view (Figures \ref{fig:overdensity2} and \ref{fig:JHK}), one of which is the radio galaxy. The spatial distribution of the JHK galaxies is presented in Figure \ref{fig:overdensity2}.

Whilst all of our \Halpha emitters are likely to lie within the protocluster, the JHK galaxies lie within a much larger redshift range and so it is unclear whether they are associated with the protocluster. Two JHK galaxies, in addition to the radio galaxy, are \Halpha emitters, meaning these galaxies are highly dust obscured, star forming galaxies which lie in the protocluster. One of these is the \Halpha source with a $3\sigma$ signal at 24\,\micron and $2\sigma$ signal at 250\,$\mu$m. Stacking the NB images for the remaining 7 JHK galaxies does not produce a signal, giving an upper limit of $SFR \sim 5.5$\,M$_{\odot}$\,yr$^{-1}$, and there is no significant detection ($<2\sigma$) in the stacked MIPS 24\,\micron and \emph{Herschel} 250\,\micron images. Hence if the remaining 7 JHK galaxies are in the protocluster the lack of NB emission indicates that they are passive, with a sSFR of $\text{log}_{10}(sSFR/\text{yr}^{-1}) \leq -9.7$.

\begin{figure*} %figure* to break the columns and put figure across whole page
\centering
%Trim option's parameter order: left bottom right top
\includegraphics[width=2\columnwidth]{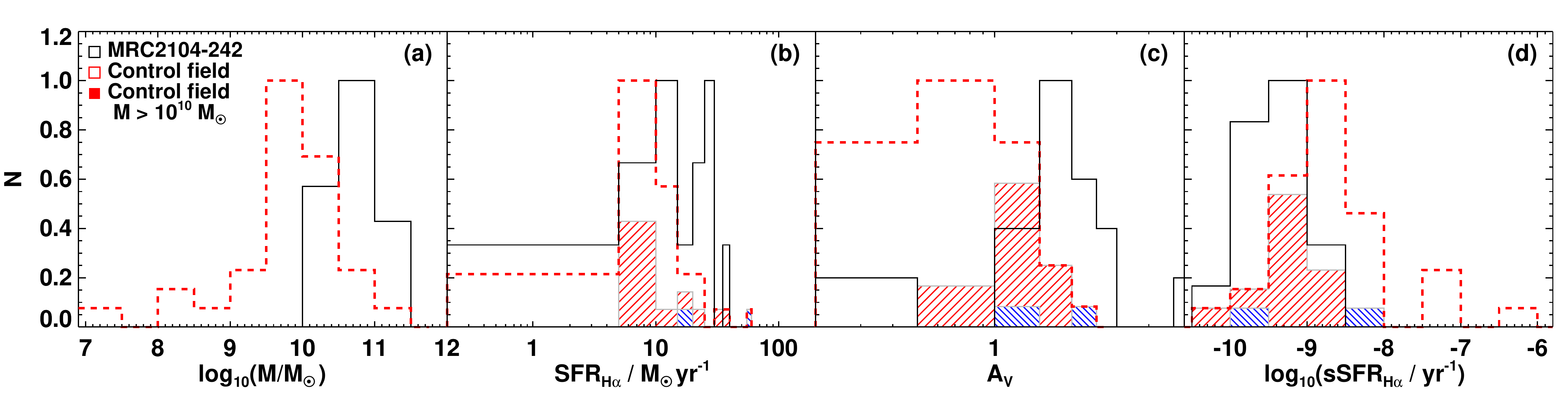}

\caption{A comparison of the properties of the protocluster galaxies (black) and control galaxies (red dashed lines) including: $(a)$ Mass, $(b)$ \Halpha SFR (dust corrected), $(c)$ $A_V$, $(d)$ sSFR. Shaded in red are the mass-selected control field histograms for SFR, $A_V$ and sSFR (log($M$/M$_{\odot}$) $> 10$). Blue shaded histograms show the two AGN candidates. Each histogram is normalised to 1. 
}
\label{fig:properties}
\end{figure*}

\subsection{Comparison of the \Halpha emitters in the protocluster and control fields}
In this section we perform a detailed comparison of the protocluster and control galaxies, including their stellar masses, SFRs, dust extinction ($A_V$), and sSFRs. In all following analysis the radio galaxy and three companions (see Figure \ref{fig:zoom}) have been removed from the protocluster sample  
as these objects are likely to be affected by the radio jets. %In the following subsections we examine the galaxy properties in more detail. 

\subsubsection{Mass} \label{sec:properties_mass}
The protocluster galaxies are on average more massive than the control field galaxies as shown by Figure \ref{fig:properties}a. A two-sided Kolmogorov-Smirnov (K-S) test shows a significant difference between the two samples: K-S $p$ $= 2.2$ $\times 10^{-5}$. The SED fits at masses $M < 10^{9} M_{\odot}$ have large errors associated with them, but even if we exclude these galaxies from our analysis there is still a significant difference (K-S\,$p = 1.1 \times 10^{-4}$). A similar difference between the masses of protocluster and control  galaxies has been found in other $z>2$ studies, including \citet{Steidel2005,Hatch2011b} and \citet{Koyama2013a}. 

The protocluster contains a large number of $M > 10^{10.5} M_{\odot}$ objects and no objects with $M < 10^{10} M_{\odot}$ within our 7\,sq.\,arcmin field-of-view. Our detection method selects on \Halpha equivalent width and galaxies below our completeness limit in SFR ($<7$\,M$_\odot$\,yr$^{-1}$) may not be selected. The \Halpha sample is therefore incomplete at all masses and particularly at low masses due to the mass-SFR relation. However we emphasise that both the protocluster and the control fields are incomplete to the same level as we have ensured that the selection method is identical in all fields.  Hence the difference in mass functions in different environments is puzzling and is discussed in detail in Section \ref{sec:lack_low_m}.

\subsubsection{Dust} \label{sec:properties_dust}
The protocluster galaxies typically have higher dust extinction, as calculated from their UV slopes, than the field galaxies, with a median $A_{V}$ that is twice as large (see Figure \ref{fig:properties}c). A K-S test shows a significant difference in the dust content between the two environments: K-S\,$p = 3.2 \times 10^{-6}$.

\begin{figure}
\centering
%Trim option's parameter order: left bottom right top
% \includegraphics[page=2, trim = 6mm 0mm 25mm 155mm, clip, scale=0.5]{SFR_mass_errs_newlim.pdf}
\includegraphics[width=\columnwidth]{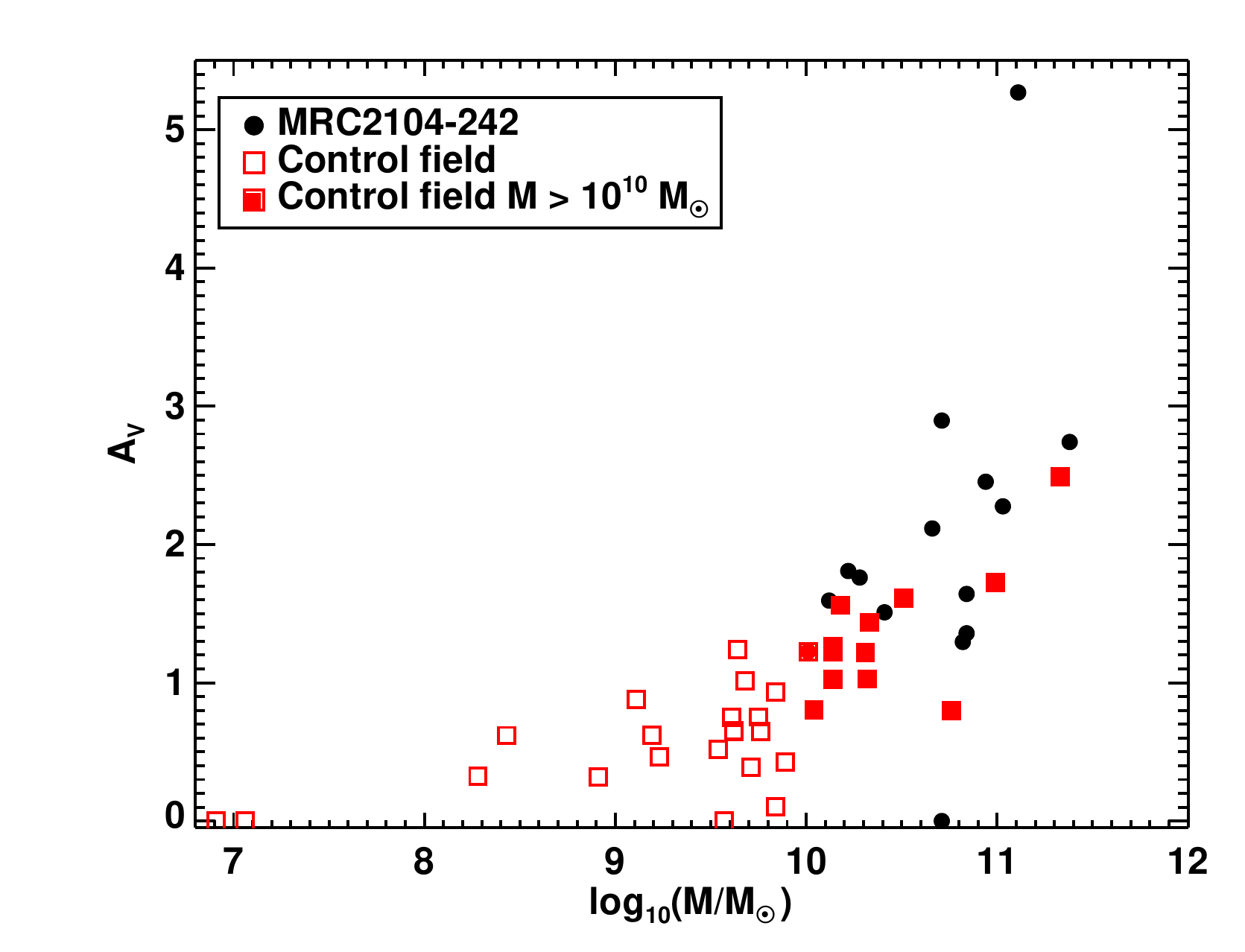}
\caption{$A_{V}$ as a function of galaxy mass for the protocluster (black circles) and control field (red squares) galaxies. %Filled red squares show the control field galaxies with $M \geq 10^{10}$\,M$_{\odot}$. 
There is a trend for increasing dust extinction with galaxy mass, with the trend becoming steeper at higher masses.}
\label{fig:AV_mass}
\end{figure}

Dust extinction correlates strongly with galaxy mass \citep[e.g.][]{Garn2010} so we tested whether the observed trend was a symptom of the mass difference found in Section \ref{sec:properties_mass} by limiting our analysis to galaxies with $M \geq 10^{10}$\,M$_{\odot}$. In Figure \ref{fig:AV_mass} we show the values of $A_{V}$ in both the protocluster and control fields as a function of mass, with filled red squares highlighting the control field galaxies with $M \geq 10^{10}$\,M$_{\odot}$. The range of $A_{V}$ reduces for this mass-limited sample and the control field galaxies are more consistent with those in the protocluster. There remains a significant difference in the dust extinction measured in the protocluster and control galaxies for this sample, however only at a 2$\sigma$ level (K-S\,$p = 0.02$).

\subsubsection{SFRs}
\begin{figure}
\centering
%Trim option's parameter order: left bottom right top
% \includegraphics[page=2, trim = 6mm 0mm 25mm 155mm, clip, scale=0.5]{SFR_mass_errs_newlim.pdf}
\includegraphics[scale=0.5]{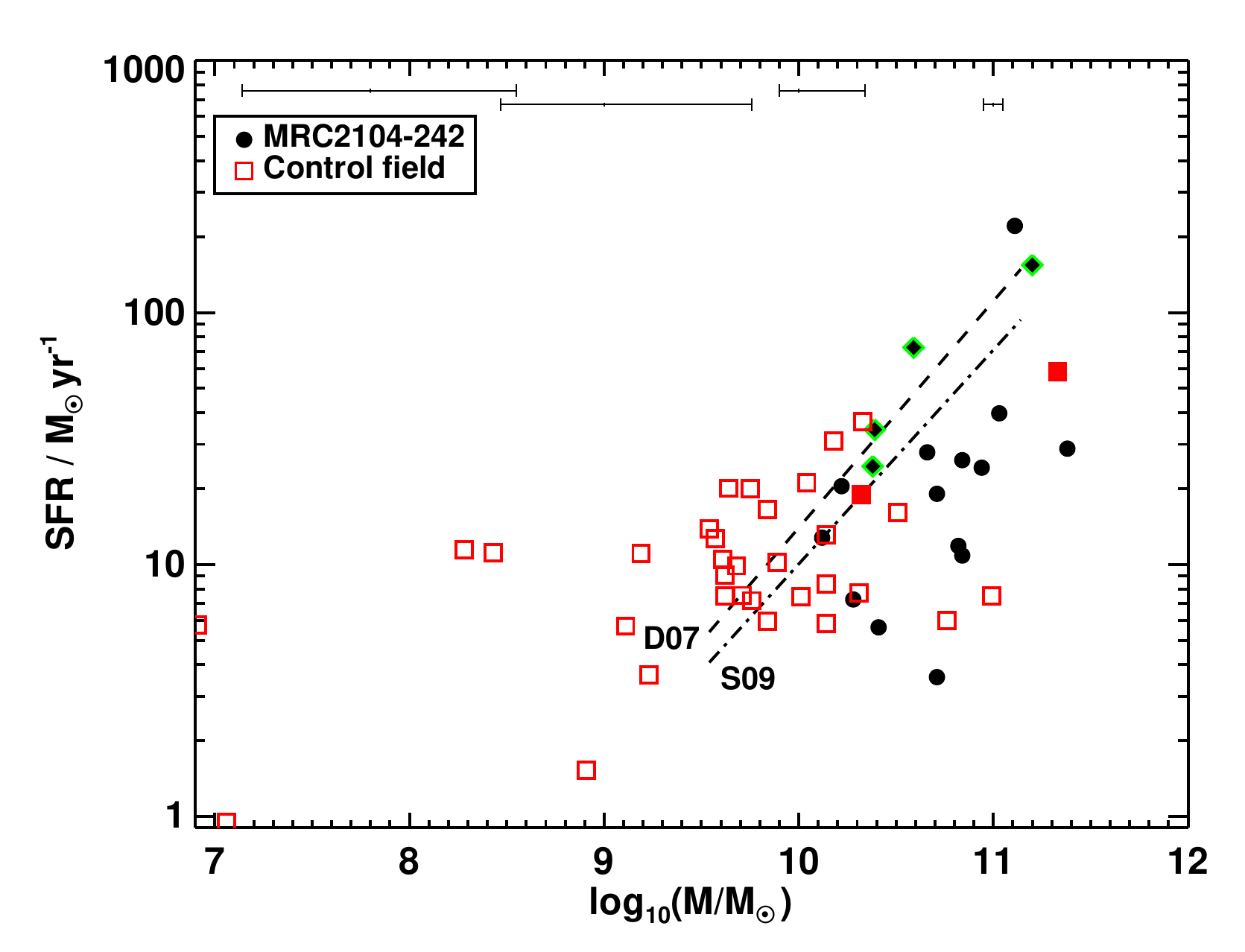}
\caption{Dust corrected \Halpha SFRs against stellar mass. The protocluster galaxies are plotted as filled, black circles and lie mostly at the high mass end of the plot. The control fields are plotted as open, red squares. The radio galaxy and three companion galaxies are highlighted with green diamonds. The COSMOS AGN candidates are plotted as filled red squares. Median error bars at the top show the typical uncertainty on the SED mass estimates in four mass bins (6.5 $\leq$ log$_{10}(M/\text{M}_{\odot})$ $< 8.5$, 8.5 $\leq$ log$_{10}(M/\text{M}_{\odot})$ $< 9.5$, 9.5 $\leq$ log$_{10}(M/\text{M}_{\odot})$ $< 10.5$, 10.5 $\leq$ log$_{10}(M/\text{M}_{\odot})$ $< 12$). Overplotted are relations from previous studies, \citep[labelled D07;S09]{Daddi2007,Santini2009}, valid above $M \sim 10^{9.5}$ M$_{\odot}$.}
\label{fig:SFRvM}
\end{figure}

The H$\alpha$ SFRs corrected for dust extinction using the UV slope are plotted in Figures \ref{fig:properties}b and \ref{fig:SFRvM}; there is little difference between the protocluster and control galaxies.  A K-S test results in a probability of $0.1$.

%We now compare the derived galaxy properties in the protocluster to the field. 
Plotted in Figure \ref{fig:SFRvM} are the dust-corrected H$\alpha$ SFRs against the SED-derived stellar masses for both the protocluster and control field galaxies.  The \citet{Daddi2007} and \citet{Santini2009} correlations showing the ``main-sequence" for $z \sim 2$ galaxies are also plotted for comparison. 
The scatter of H$\alpha$ emitters with $M<10^{10.5}$\,M$_{\odot}$ is consistent with the main sequence, but at higher masses both the protocluster and control field galaxies appear to lie below this relation. This suggests that the applied dust-correction for the high-mass H$\alpha$ emitters is not sufficient and there may be additional star formation that is heavily optically obscured. It is extremely difficult to correct for dust extinction using the UV slope alone \citep{Elbaz2011} and a far more accurate measurement of the total SFR is obtained through the IR luminosity.

In Figure \ref{fig:MIPS} we show the total SFR derived by combining the raw \Halpha SFRs with SFRs derived through the IR 24\,\micron and 250\,\micron luminosities. SFRs derived using 24$\mu$m have two values depending on whether we assume they have ULIRG SEDs or whether they have main-sequence SEDs. 
We note that as \Halpha emission is less sensitive to dust attenuation than rest-frame UV light, these total SFRs may slightly overestimate the true SFR. However, the derived total SFRs are almost entirely dominated by the IR, so the contribution from unobscured \Halpha is likely to be negligible. 
The \emph{Herschel} 250\,\micron protocluster SFR estimate is in better agreement with the 24\,\micron IR SFR estimate based on local ULIRGs \citep{Rieke2009},  although all of these IR estimates are in agreement with the main sequence relationship.  Whilst the 24\,\micron signal could be due to AGN-heated warm dust, the detection of 250\,\micron flux (rest-frame 70\,$\mu$m) in the protocluster galaxies indicates that we must be detecting cooler dust heated by UV emission from young, hot stars. 

The IR$+$\Halpha SFRs are comparable to the dust-corrected \Halpha SFRs in the control fields, but in the protocluster we find a large discrepancy. The IR$+$\Halpha SFRs are at least twice as fast (and up to ten times as fast) as the dust-corrected \Halpha SFRs which implies the protocluster galaxies contain more optically-obscured star formation than in the control galaxies. These results imply that the total SFR of the massive galaxies which reside in dense regions cannot be derived from \Halpha estimates alone; the protocluster galaxies have higher masses with large dust extinctions, therefore far-IR or sub-millimetre data are required to probe the optically-obscured star formation. We note that the large amount of dust extinction may have implications for studies which aim to detect protoclusters and study them through Lyman $\alpha$ emission from their member galaxies.

The IR SFRs reveal a different picture to the \Halpha SFRs: the protocluster galaxies are forming stars more rapidly than the control galaxies but much of this star formation is hidden from optical view. Figure \ref{fig:MIPS} reveals that once this obscured star formation is taken into account the protocluster galaxies lie on the same main sequence of the mass-SFR relation as the control galaxies. 

Our  IR SFR estimates for both control and protocluster galaxies are consistent with the main-sequence of the SFR-mass relation, suggesting that the majority of these \Halpha emitters are not undergoing a ``bursty" mode of star formation but rather forming stars at the expected rate for their mass. This is in agreement with previous protocluster studies \citep{Koyama2013a,Koyama2013b}. However, we note that the SFR$_{IR}$ are derived from median stacks, thus our method would not be able to find starbursting galaxies if the majority of the \Halpha emitters were main sequence galaxies. A few of the protocluster galaxies have 2$\sigma$ detections at 250\,$\mu$m, and one has a 3$\sigma$ detection at 24\,$\mu$m. 
If we remove from the 250\,\micron stack those \Halpha emitters with nearby ($\le 10$\,arcsec) $2\sigma$ detections, the signal of the stack decreases and we do not find a signal above 3$\sigma$ (where 3$\sigma$ corresponds to an upper limit of 98\,M$_{\odot}$\,yr$^{-1}$). We discuss these galaxies further in Section \ref{sec:starbursts}.

\subsubsection{sSFRs}
Figure \ref{fig:properties}d compares the specific star formation rates (sSFR) of the protocluster and control galaxies. When the entire mass range of galaxies is taken into account there is a significant difference in the sSFRs between the two populations (K-S\,$p = 6.8 \times 10^{-4}$). However this difference is driven by the disparate mass distributions of galaxies in the two environments. 
The shaded red histogram shows the distribution of sSFRs of galaxies with masses $M \geq 10^{10}$\,M$_{\odot}$. For this population there is no significant difference in the sSFRs: K-S $p = 0.15$.

\subsection{Highly starforming galaxies} \label{sec:starbursts}
\begin{figure}
\centering
%Trim option's parameter order: left bottom right top
% \includegraphics[trim = 5mm 0mm 20mm 155mm, clip, scale=0.5]{SFR_IR_mass.pdf}
\includegraphics[scale=0.5]{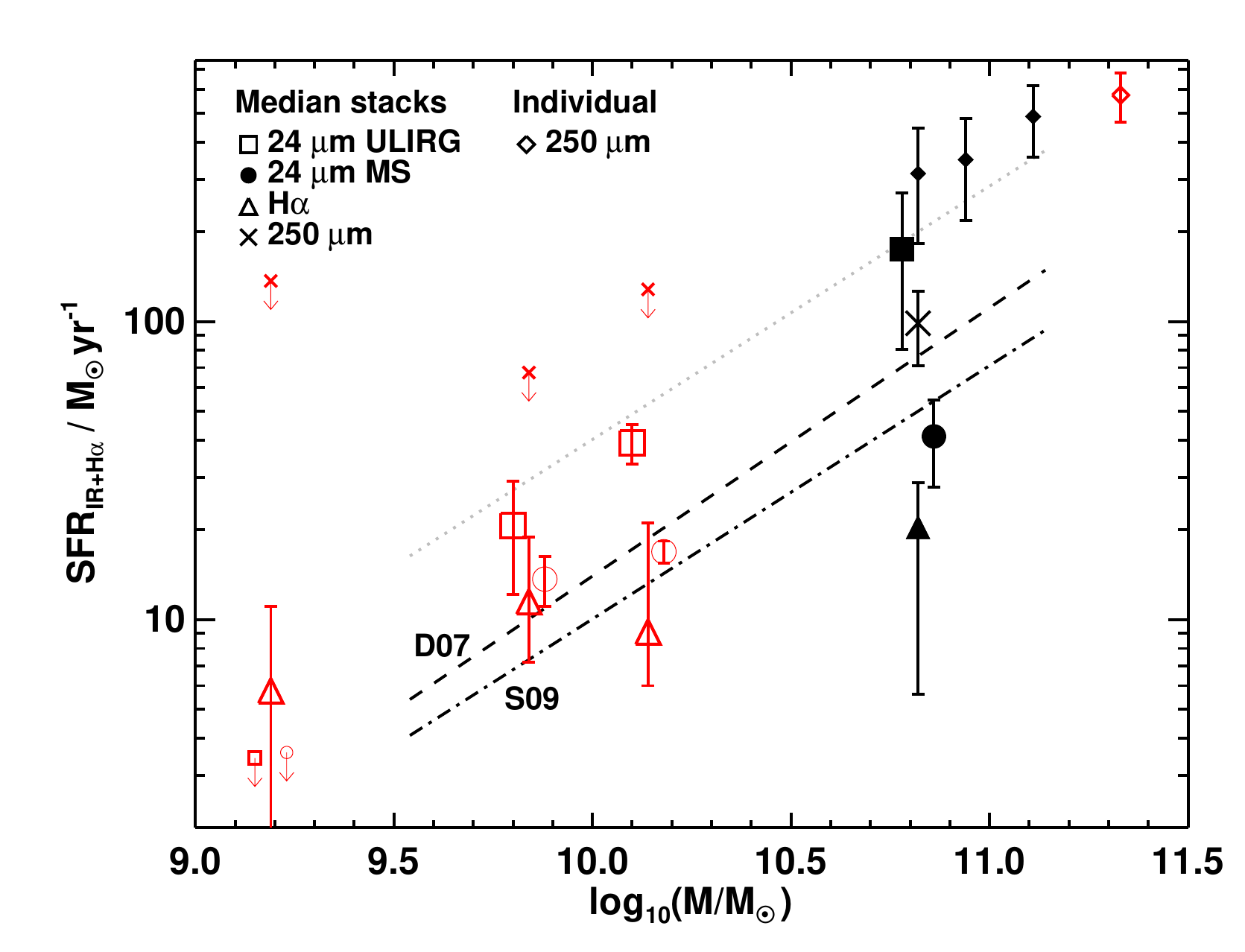}
\caption{The IR$+$H$\alpha$ SFRs plotted against median mass values for the protocluster (black filled symbols) and control fields (red open symbols). From left to right: GOODS-S, COSMOS, UDS. %The GOODS-S 24\,\micron stack showed no signal and so a 
$3\sigma$ upper limits are plotted if no signal is observed. Circles and squares (offset in the $x$-axis for clarity) are SFRs calculated from 24\,\micron using \citet{Rujopakarn2013} and \citet{Rieke2009} respectively. Triangles are the dust-corrected \Halpha SFRs, with error bars enclosing 68\% of the data. 
The crosses are the SFRs derived from \emph{Herschel} 250\,\micron stacks of \Halpha emitters.  
Overplotted are \citet{Daddi2007} and \citet{Santini2009} relations, labelled D07 and S09.  Also plotted are the \emph{Herschel} SFRs (IR$+$H$\alpha$) and 1$\sigma$ error bars for sources with 250\,\micron signal $> 2\sigma$ above the background noise: black/red diamonds are protocluster/COSMOS galaxies. These sources are consistent with being starbursts ($\sim 4 \times$ the main sequence); the grey dotted line shows the \citet{Santini2009} relation multiplied by 4.}
\label{fig:MIPS}
\end{figure}

No \Halpha emitters in the protocluster or control fields are detected above $3\sigma$ significance at 250\,$\mu$m, however there are a few detections with signals $>2\sigma$. In the MRC\,2104$-$242 field there are three 2$\sigma$ sources, one of which has a $3\sigma$ 24\,\micron detection of $0.11$\,$\mu$Jy$ = 145\pm60$\,M$_{\odot}$\,yr$^{-1}$ (MS) or $= 1200 \pm 775$\,M$_{\odot}$\,yr$^{-1}$ (ULIRG). Their 250\,\micron SFRs are plotted in Figure \ref{fig:MIPS} as small black diamonds.

In the control fields we only find one source with a $>$\,2$\sigma$ detection. The 250\,$\mu$m-derived SFR is plotted as a small red diamond in Figure \ref{fig:MIPS}. This source is one of the AGN candidates in the COSMOS field.

We expect 5\% of our \Halpha emitters (i.e. $<1$ of the \Halpha emitters) to be detected at the 2$\sigma$ level due to noise in the 250\,\micron data. In the protocluster we find three, suggesting that at least two of them are real sources and not noise. All three sources have 250\,\micron SFRs which are consistent with starbursting galaxies, defined such that they lie four times above the main sequence \citep{Rodighiero2011}. This suggests that the fraction of starbursts is several times higher in the protocluster, with 21\% of the \Halpha emitters being starburst galaxies, compared to just $\sim3$\% in the control field.

\subsection{Robustness checks}
\subsubsection{AGN} \label{sec:robust_agn}
Removing the two AGN detected in the COSMOS field from our control sample does not significantly change our results. There is still a significant difference in dust content estimated from the UV slope (K-S\,$p = 2 \times 10^{-6}$) which remains at a 2$\sigma$ level when considering the mass-limited galaxy samples. Furthermore the trends for the sSFRs remain the same: K-S\,$p = 2.5 \times 10^{-4}$ and K-S\,$p = 0.1$ for the full sample and mass-limited sample respectively.  The average  IR and \Halpha SFRs decrease for the COSMOS field\footnote{The median dust-corrected H$\alpha$ SFR for the COSMOS field decreases by $< 1$\,M$_{\odot}$\,yr$^{-1}$ and the 24\,\micron + H$\alpha$ SFR decreases by $\sim 2$\,M$_{\odot}$\,yr$^{-1}$.} and the \Halpha SFR distributions become significantly different at a $2\sigma$ level (K-S\,$p$ $=$ $0.05$). However, in the mass-limited sample (M$>10^{10}$\,M$_{\odot}$) there is still no significant difference in the SFRs between the two distributions: K-S\,$p$ $=$ $0.43$. Excluding the COSMOS AGN, the starburst galaxy fraction is still higher in the protocluster than the control field.

\subsubsection{Luminosity distances}
The NB filters used for our control fields have different central wavelengths from the NB229 filter used to select the protocluster galaxies. Since we select galaxies at slightly different redshifts, the luminosity distance to the control field galaxies is slightly less than to the protocluster galaxies. As the control field galaxies are at lower redshifts than the protocluster,  we probe further down the luminosity function of the control field for the same cuts in apparent magnitude. We have tested how this may affect the results by taking this difference in magnitude into account and applying a cut to the control fields at brighter magnitudes. These cuts remove five control field galaxies from our sample, increasing the level of overdensity measured in the protocluster field to $9\pm 0.8$ times the control field density. The masses and star formation properties of the remaining galaxies remain within the error margins calculated. So the difference in luminosity distance between the protocluster and control fields does not affect our other conclusions.

\section{Discussion} \label{sec:discussion}

\subsection{Galaxy growth in protoclusters}
We have shown that the star forming protocluster galaxies at  $z = 2.5$ are more massive than similarly selected galaxies in the field. The SFRs and sSFRs of the protocluster galaxies are consistent with the control galaxies once we take into account the difference in galaxy mass by only comparing galaxy samples of similar mass.

The high-mass protocluster galaxies include a larger amount of dust-obscured star formation than the lower-mass control galaxies.  Once this has been included in the total SFRs by adding the IR SFRs from the 24\,\micron and \emph{Herschel} 250\,\micron data, we find that on average, protocluster and control galaxies lie on the same main-sequence of the SFR-mass relation. 
This means that at $z \sim 2.5$, galaxy growth in terms of star formation is regulated predominantly by galaxy mass and is not greatly affected by the environment of the host galaxy. 

Figures \ref{fig:properties}a and \ref{fig:SFRvM} show that the protocluster galaxies typically have higher masses than the control field galaxies, and there are more than twice as many protocluster galaxies than field galaxies with $M > 10^{10.5}$\,M$_{\odot}$ (10 protocluster galaxies compared to only 4 control field galaxies even though the protocluster area surveyed is only 4\% of the control area). This poses a conundrum: 
if the SFR is governed by galaxy mass alone at $z\sim2.5$, then how did the protocluster galaxies gain so much mass so rapidly? The early formation of these galaxies must be dependent on their environments at higher redshift, even though at $z \sim 2.5$ their growth proceeds in the same way.

We find three 2$\sigma$ detections at 250\,\micron in the protocluster, suggesting the presence of starbursting galaxies. If the fraction of galaxies undergoing a starburst is much greater in denser environments, this may explain the higher masses. Deeper submm observations of protocluster galaxies are essential to understanding this issue. 

\begin{figure}
\centering
%Trim option's parameter order: left bottom right top
% \includegraphics[trim = 5mm 0mm 20mm 155mm, clip, scale=0.5]{Schechter_newlim.pdf}
\includegraphics[width=\columnwidth]{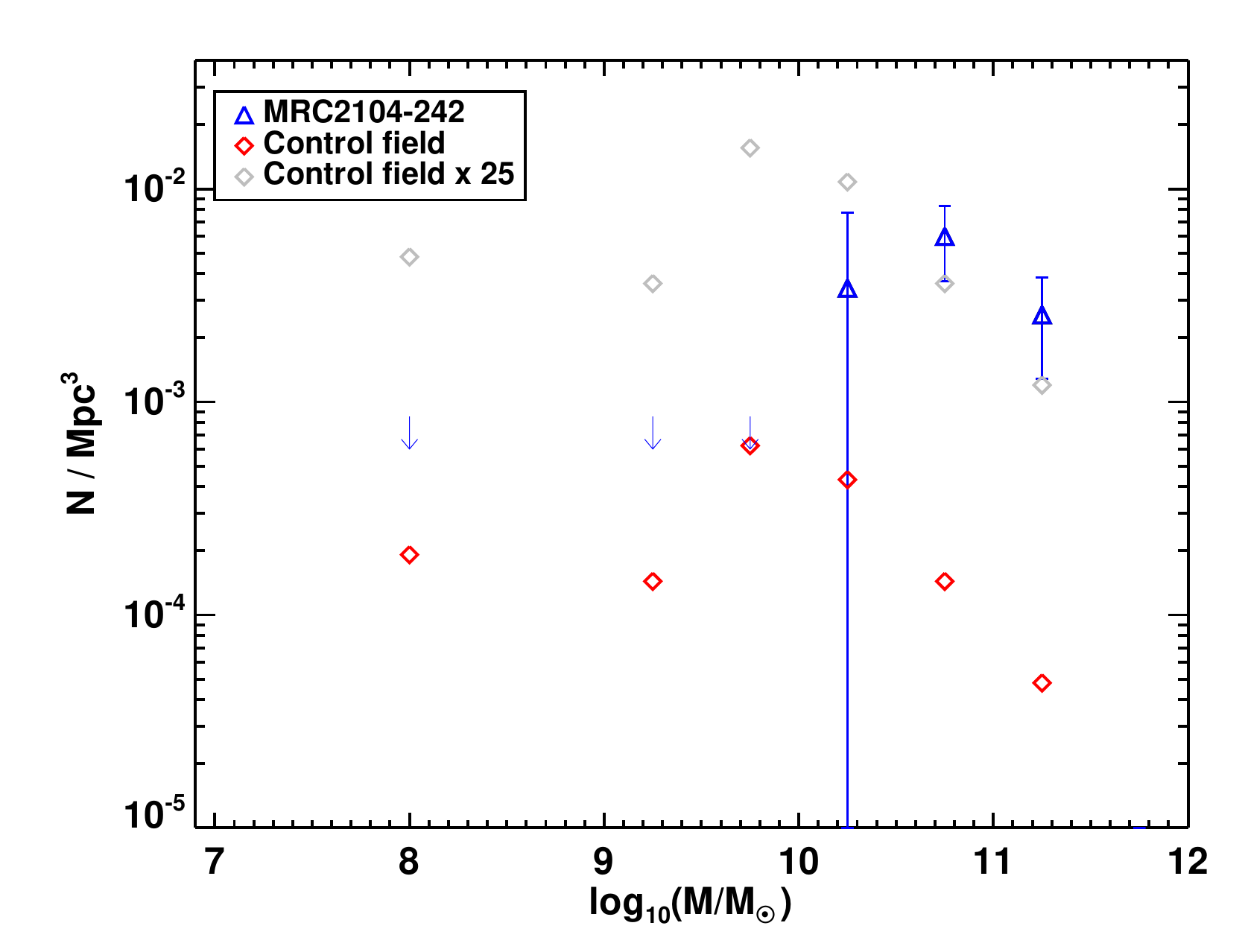}
\caption{Galaxy number densities per mass bin for the control field (red diamonds) and the protocluster (blue triangles). In grey we also show the control field distribution, scaled by a factor of 25, to illustrate the expected number densities in the protocluster. This figure shows a clear excess of galaxies in the protocluster at the high mass end, however there appears to be a lack of low mass objects in the protocluster, whereas we detect many low mass objects in the field.
}
\label{fig:Schechter}
\end{figure}

\begin{figure}
\centering
%Trim option's parameter order: left bottom right top
% \includegraphics[trim = 5mm 0mm 20mm 155mm, clip, scale=0.5]{Schechter_newlim.pdf}
\includegraphics[width=\columnwidth]{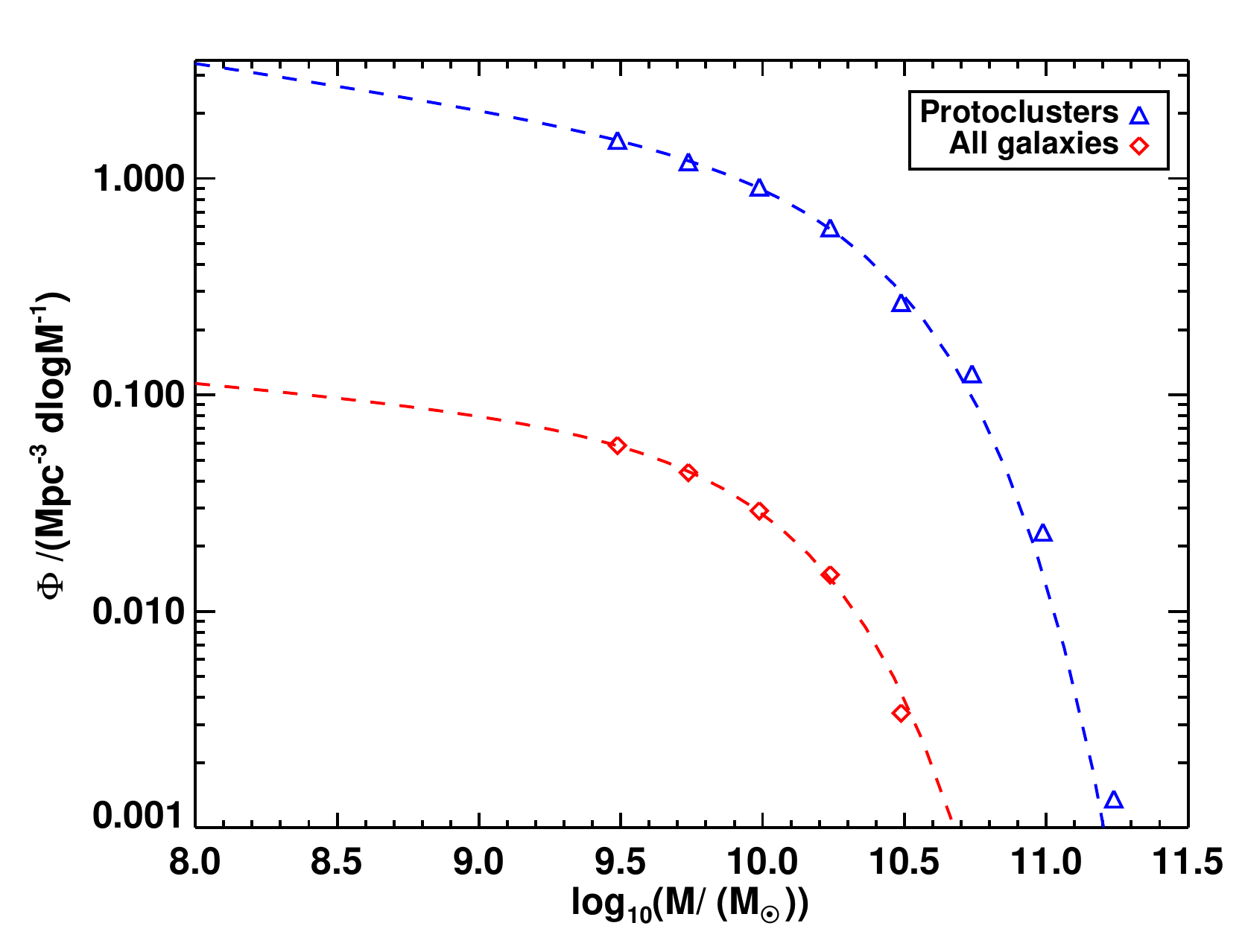}
\caption{Semi-analytic derived mass distributions for all galaxies (red diamonds) and protocluster galaxies (blue triangles, see text for details). Dashed lines show the fitted Schechter function. The values of $M_*$ and $\Phi$ differ by $0.4$\,dex and $0.14$\,dex respectively. The difference in $\alpha$ is $0.08$\,dex. }
\label{fig:Schechtermods}
\end{figure}

\subsection{Overdensity and the lack of low mass star forming galaxies in protocluster} \label{sec:lack_low_m}
%In addition, we find no low mass (M $< 10^{10} M_{\odot}$) galaxies in our protocluster sample. 
We now examine why, on average, galaxy masses differ between the two environments. We find no difference at the high mass end of the distributions; taking a mass selected sample of all \Halpha emitters with $M \geq 10^{10}$\,M$_{\odot}$ there is no significant difference in the mass distributions. However, we find no low mass ($M < 10^{10}$\,M$_{\odot}$) galaxies in our protocluster sample. 
This skew in the mass distribution means that the strength of the overdensity that we detect depends on the mass range we examine, e.g. the protocluster number density is $\sim 25$ times the control field if we only consider objects with $M > 10^{10}$\,M$_{\odot}$ and $\sim 55$ times the control field at $M > 10^{10.5}$\,M$_{\odot}$ (see Figure \ref{fig:Schechter}). 
This large excess of high-mass galaxies suggests the presence of a galaxy protocluster, as discussed in Section \ref{sec:overdensity}. If the MRC\,2104$-$242 field does contain a protocluster then we also expect to find an overdensity of low mass galaxies within the field. 
Although we are incomplete in mass, particularly at low masses, we are incomplete to the same level in the protocluster and the control fields. Since we detect 22 \Halpha emitters at $M < 10^{10}$\,M$_{\odot}$ in the control fields, we expect to detect $\sim 21$-$22$ \Halpha emitters in the protocluster, assuming an overdensity of 24, whereas we do not detect any (Figure \ref{fig:properties}a). 
We note that the \citet{Koyama2013a} study shows that the protocluster around MRC\,1138$-$262 (the Spiderweb galaxy) also lacks low-mass objects. The difference we find in the average masses between the MRC\,2104$-$242 field and the control field is due to this lack of low mass galaxies in the protocluster, rather than a population of extremely massive galaxies. 

In the following subsections, we consider three possible reasons for this difference in the protocluster and control field mass distributions: an intrinsic difference in mass functions between the protocluster and the field galaxies; observational effects, such as the higher value of dust extinction in protocluster galaxies  
or low mass galaxies which may have already shut down their star formation; and mass segregation, with high mass galaxies preferentially clustered around the radio galaxy.

\subsubsection{Environmental dependence of the galaxy mass functions}
In order to determine an expected mass function for protoclusters at $z \sim 2.5$, we use semi-analytic models to produce the mass distributions of a protocluster and the surrounding field. 
%STUART'S BIT ABOUT THE MODELS STARTS HERE
We have taken the $z=2.42$ output of the \citet{Guo2011} semi-analytic model built upon the Millennium Dark Matter Simulation \citep{Springel2005}.  The Millennium Simulation follows the evolution of $2160^3$ dark matter particles from $z=127$ to the present day in a box of comoving side length $500\,h^{-1}{\rm Mpc}$.  The simulation adopts a flat $\Lambda$CDM cosmology with $\{\Omega_0,\Omega_\Lambda,\sigma_8,n,h\}=\{0.25,0.75,0.9,1,0.73\}$.  This is consistent with the Two-Degree Field Galaxy Redshift Survey \citep[2dFGRS;][]{Colless2001} and the Wilkinson Microwave Anisotropy Probe first year results \citep[WMAP;][]{Spergel2003}, but is marginally discrepant with the latest measurements of cosmological parameters \citep{Planck2013}.  Haloes were identified using a Friends-of-Friends algorithm \citep[FoF;][]{Davis1985} with linking length 0.2, which were then analysed for bound substructures using \textsc{subfind} \citep{Springel2001}.  Only haloes containing 20 particles were considered and we note that similar results can be found with other halo finders \citep{Muldrew2011,Knebe2011}.

Galaxies were added to the halo merger tree using the \citet{Guo2011} semi-analytic model, which is an updated version of the \citet{Croton2006} and \citet{DeLucia2007} models.  A full description of the model, including modifications, can be found in those papers.  Traditionally semi-analytic models have been poor at reproducing the redshift evolution of the galaxy stellar mass function.  As shown in figure 23 of \citet{Guo2011}, the high mass end of the galaxy stellar mass function is reproduced well in this redshift range, but there is an over-abundance of lower mass galaxies.  In order to minimise the effect of this over-abundance of low mass galaxies, we limit our sample to galaxies with stellar masses greater than $10^{9.5}\,h^{-1}{\rm M_{\odot}}$.

We identify $1938$ clusters in the $z=0$ catalogue by considering haloes with masses greater $10^{14}\,h^{-1}{\rm M_{\odot}}$.  For each $z=0$ identified cluster, we locate the highest mass progenitor galaxy in the $z=2.42$ catalogue.  We then subsample cubes of side length $3.5\,h^{-1}{\rm Mpc}$ comoving centred on these progenitors and compare the mass function with that of the whole volume, a cube of side length $500\,h^{-1}{\rm Mpc}$ comoving.

%ENDS HERE

Fitting a Schechter curve to the semi-analytic derived mass distributions (Figure \ref{fig:Schechtermods}) we find that the expected mass function of the protocluster shows no turnover at the faint end. Indeed, we find the faint end slope for protocluster galaxies tends to be slightly steeper (by ~0.1 dex) than that for the whole volume. The distributions differ significantly in the value of $M_*$ and normalisation (differences of $0.4$\,dex and $0.14$\,dex respectively). This means that the difference in number densities that we observe is not due to a fundamental difference in the shape of the mass functions at $z \sim 2.5$. 
The shape of the expected mass function is dependent on the volume sampled and the SFR of the galaxies that are selected. We will further examine how the star forming fraction, and hence galaxy mass function, changes as a function of volume sampled in an upcoming paper (Muldrew et al. in prep). 

\begin{figure}
\centering
%Trim option's parameter order: left bottom right top
% \includegraphics[trim = 5mm 0mm 20mm 155mm, clip, scale=0.5]{Schechter_newlim.pdf}
\includegraphics[width=\columnwidth]{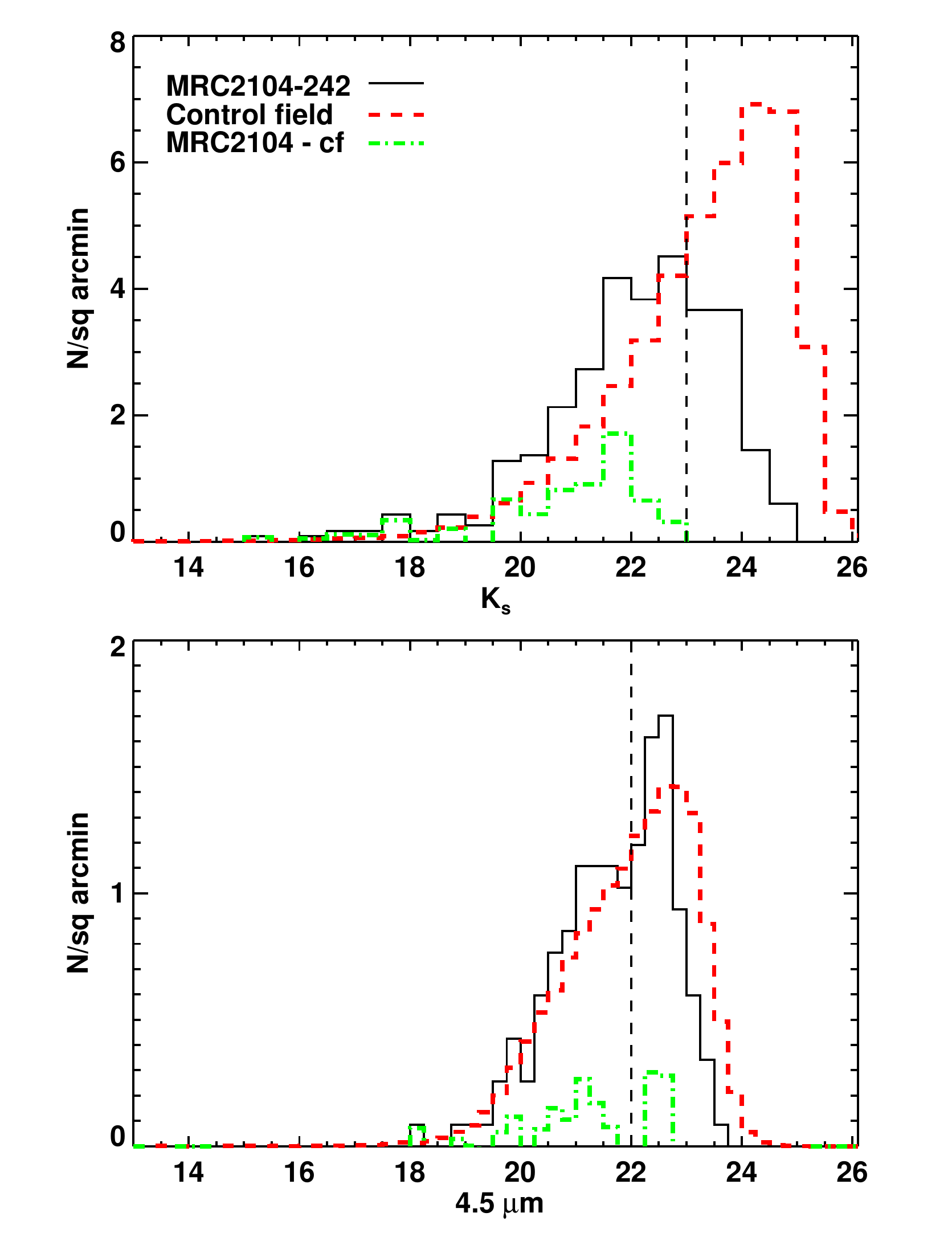}
\caption{Number density histograms in the $K_s$ band and at 4.5\,$\mu$m for galaxies with colours $J-H > H-K_s-0.15$ (Vega) and $[3.6]-[4.5] > -0.1$ (AB) respectively. These criteria select passive, as well as star-forming, galaxies. Black histograms are for the MRC\,2104$-$242 field, red is the control field and green is the difference between the two, indicating protocluster candidates. Completeness is shown by the vertical dashed lines. The lack of protocluster galaxies %compared to the control field 
at magnitudes brighter than the completeness limits, shown by the drop in the green histograms, suggests a lack of faint galaxies in this protocluster.}
\label{fig:lfns}
\end{figure}

\subsubsection{Observational effects on mass distributions}
%In this subsection we examine whether our results could be an observational effect. 
Our NB survey selects star forming galaxies with \Halpha emission, down to a dust-uncorrected star formation rate of $\sim 7$\,M$_{\odot}$\,yr$^{-1}$. If the low mass protocluster galaxies were passive or heavily obscured by dust, our NB survey would not detect them. 
To test if these galaxies are missing in our NB survey, we compare the galaxy luminosity functions in the protocluster field to the control field (Figure \ref{fig:lfns}). 

We compare the luminosity functions in the $K_s$ band, using a $J-H > H-K_s-0.15$ cut to remove galaxies at redshifts below $\sim1$, and at 4.5\,$\mu$m, taking a $([3.6]-[4.5])_{AB} > -0.1$ colour cut (selecting galaxies at $z>1.4$). These wavebands select passive galaxies, as well as the star-forming NB emitters in our sample, albeit with a large contamination rate. 
We find an overdensity of bright galaxies ($K_s < 21.9$ and 4.5\,$\mu$m $< 20.5$) and a lack of faint galaxies in both $K_s$ and 4.5\,$\mu$m at magnitudes fainter than 21.9 (AB) and 20.5 (AB) respectively. The lack of faint galaxies, at magnitudes brighter than the completeness limits (shown by the vertical dashed lines in Figure \ref{fig:lfns}), suggests that this protocluster lacks both star forming and passive low mass galaxies. 

Recently \citet{Kulas2013} found that the metallicity of protocluster galaxies did not vary with galaxy mass, whereas field galaxy metallicity decreases with decreasing mass. They found no difference between the two environments at high masses, but at low masses found a significant difference in metallicity. This suggests that low mass galaxies are more metal rich in protocluster environments than in the field. This may also mean that the low mass galaxies in protoclusters are dustier than those in the field. 
However, with our current data we find no evidence to suggest this and it is difficult to test as we do not detect any low mass galaxies in the protocluster.

\subsubsection{Mass segregation}
Protoclusters at high redshift are not dynamically evolved and so it is unlikely that large-scale mass segregation has had enough time to occur: our 2.65\,arcmin $\times$ 2.65\,arcmin area corresponds to 1.28\,Mpc $\times 1.28$\,Mpc in physical coordinates. Assuming an average galaxy velocity of $500 \text{\,km\,s}^{-1}$, this gives a crossing time of $2.5$\,Gyr. At $z=2.49$, the age of the Universe is $2.58$\,Gyr. This means that there has not been enough time for virialization to occur and any dynamical friction effects will not be strong enough to produce mass segregation in the protocluster at this redshift. 

Substructure has, however, been found around radio galaxies at high redshift. \citet{Hayashi2012} reported the discovery of a protocluster where there were three distinct ``clumps" of galaxies on scales of $\sim 8-10$\,Mpc. They found that the highest mass objects resided in the densest clump at $z = 2.53$, suggesting that higher mass objects may preferentially form in denser environments. \citet{Kuiper2010} also found that the most massive and highly star forming galaxies were located near the radio galaxy of a $z \sim 3$ protocluster. 
It may be that protoclusters have more high mass galaxies forming through monolithic collapse, or experience many more mergers in the early years of their formation. Measuring galaxy sizes in protoclusters compared to the field may provide more information on galaxy formation mechanisms in different environments. 

\subsubsection{Where are the low mass galaxies?}
In the previous subsections we have established that the MRC\,2104$-$242 protocluster galaxy mass function differs from that of the control field for both star forming and passive galaxies. 
A higher level of dust extinction in only the low mass protocluster galaxies could produce this effect observationally; with our current data we only find a 2$\sigma$ difference in the dust extinction between the protocluster and control field galaxies at high masses, and cannot test this at lower masses. Alternatively, protocluster enivronments may form more high mass galaxies through monolithic collapse or protocluster galaxies may undergo many more mergers in the early stages of their growth compared to the field. 
We find tentative evidence that the fraction of starburst galaxies is higher in the protocluster, indicating a more rapid growth of galaxies in denser environments. 
We note that data from \citet{Koyama2013a} also shows a similar lack of galaxies with low masses in the MRC\,1138$-$262 protocluster, however with only two protoclusters it is difficult to come to any firm conclusions as to why we find this result. In future studies it is important that we now progress towards larger samples of protoclusters, in order to obtain a meaningful statistical understanding of the formation and evolution of these structures.

\section{Conclusions and summary} \label{sec:summary}
We have undertaken a NB survey of the field around the HzRG MRC\,2104$-$242. We have selected star-forming galaxies in this field and compared their properties with those of a field sample at similar redshifts. Here we present our key results:
\begin{enumerate}
\parsep0pt
\item The field around the HzRG MRC\,2104$-$242 is overdense compared to blank control fields, with a level of overdensity of $8.0\pm0.8$ times the average blank field, which is consistent with this field being the progenitor of a low redshift cluster, i.e. a protocluster. 
\item The protocluster galaxies around MRC\,2104$-$242 are more massive and have more hidden star formation than control field galaxies at the same redshift. When we take a mass selected field sample we find no difference in the SFR and sSFR between the two environments, and only a minor difference in the dust content. 
\item Star formation at $z \sim 2.5$ is governed predominantly by galaxy mass, not environment. After including dust-extincted star formation using 24\,\micron and \emph{Herschel} data we find that the average SFR-mass relations are the same irrespective of environment and both the protocluster and control field galaxies lie close to the main sequence.
\item We find a large difference in the mass distributions between environments: we expect to find $\sim21$-$22$ galaxies in the protocluster at masses $M<10^{10}$\,M$_{\odot}$ and detect none. This could indicate a higher level of dust extinction in low mass galaxies in the protocluster. It may alternatively be due to galaxies in the protocluster forming more high mass galaxies through monolithic collapse or undergoing many more mergers in the early stages of their growth. 
\item We find tentative evidence of a larger fraction of starburst galaxies in the protocluster than in the control field. Further data is required to confirm the 250\,\micron detections, however a more rapid mode of star formation in denser environments may explain how protocluster galaxies build up their mass quicker than in the field. 
\item The overdensity we detect in this small area is highly dependent on the mass range we consider. It can range from an overdensity of $0$ (at $M < 10^{10}$\,M$_{\odot}$) to $55$ ($M > 10^{10.5}$\,M$_{\odot}$). 
It is important when quantifying protoclusters to compare their mass functions, rather than simply number overdensities. 
\end{enumerate}

\section*{Acknowledgments}
We thank Bruce Sibthorpe for help with the Herschel data reduction and Dan Smith for useful discussions. 
We thank the referee for their helpful suggestions which improved the paper. 
EAC acknowledges support from the STFC. NAH is supported by an STFC Rutherford Fellowship. SIM acknowledges the support of the STFC Studentship Enhancement Programme (STEP). EER acknowledges financial support from NWO (grant number: NWO-TOP LOFAR 614.001.006). 

The research in this paper is based largely on observations made with ESO Telescopes at the La Silla Paranal Observatory under programme IDs 081.A-0673 and 088.A-0954. This work also made use of observations made with the Spitzer Space Telescope, which is operated by the Jet Propulsion Laboratory, California Institute of Technology under a contract with NASA, as well as observations obtained at the Gemini Observatory, programme GS-2010B-Q-65. The Gemini Observatory is operated by the Association of Universities for Research in Astronomy, Inc., under a cooperative agreement with the NSF on behalf of the Gemini partnership: the National Science Foundation (United States), the National Research Council (Canada), CONICYT (Chile), the Australian Research Council (Australia), Minist\'{e}rio da Ci\^{e}ncia, Tecnologia e Inova\c{c}\~{a}o (Brazil) and Ministerio de Ciencia, Tecnolog\'{i}a e Innovaci\'{o}n Productiva (Argentina). The Herschel-ATLAS is a project with Herschel, which is an ESA space observatory with science instruments provided by European-led Principal Investigator consortia and with important participation from NASA. The H-ATLAS website is http://www.h-atlas.org/.

The Millennium Simulation used in this paper was carried out by the Virgo Supercomputing Consortium at the Computing Centre of the Max-Planck Society in Garching.  The halo merger trees used in the paper are publicly available through the German Astronomical Virtual Observatory (GAVO) interface, found at http://www.mpa-garching.mpg.de/millennium/.

\bibliographystyle{mn2e}
\bibliography{protoclustersbib}

\bsp

\label{lastpage}

\end{document}